\documentclass{ieeeaccess}
\usepackage{cite}
\usepackage{amsmath,amssymb,amsfonts}
\usepackage{graphicx}
\usepackage{textcomp}
\usepackage{adjustbox}
\usepackage{color,soul}
\sethlcolor{white}

\usepackage{algorithm}
\usepackage{algpseudocode}

\usepackage{caption}
\usepackage{xurl}
\usepackage{lipsum}
\usepackage{multicol}
\usepackage[utf8]{inputenc}
\usepackage{amssymb}
\usepackage{subfigure}
\usepackage{color}
\usepackage{url}
\usepackage{multirow}
\usepackage{epsfig}
\usepackage{epstopdf}
\usepackage{graphicx}
\usepackage{subfig}
\usepackage{amssymb}
\usepackage[figuresright]{rotating}
\usepackage{float}
\usepackage{fancyhdr}

\usepackage{ragged2e}

\def\BibTeX{{\rm B\kern-.05em{\sc i\kern-.025em b}\kern-.08em
    T\kern-.1667em\lower.7ex\hbox{E}\kern-.125emX}}
    
\fancypagestyle{specialfooter}{%
  \fancyhf{}
  
  \fancyfoot[R]{Some special text}
}

\begin{document}
\thispagestyle{specialfooter}%

\history{Received March 8, 2022, accepted April 4, 2022, date of publication April 11, 2022, date of current version April 15, 2022.}
\doi{10.1109/ACCESS.2022.3166152}

\title{Joint Optimisation of Privacy and Cost of in-App Mobile User Profiling and Targeted Ads}

\author{\uppercase{Imdad Ullah} and
\uppercase{Adel Binbusayyis}}
\address[1]{College of Computer Engineering and Sciences, Prince Sattam bin Abdulaziz University, Al-Kharj 11942, Saudi Arabia. (e-mail: i.ullah@psau.edu.sa; a.binbusayyis@psau.edu.sa)}

\tfootnote{This work was supported by the Deanship of Scientific Research at Prince Sattam Bin Abdulaziz University under Grant 2021/01/17717.}

\markboth
{I. Ullah, A. Binbusayyis: Joint Optimization of Privacy and Cost of in-App Mobile User Profiling and Targeted Ads}
{I. Ullah, A. Binbusayyis: Joint Optimization of Privacy and Cost of in-App Mobile User Profiling and Targeted Ads}

\corresp{Corresponding author: Imdad Ullah (i.ullah@psau.edu.sa).}

\begin{abstract}
Online mobile advertising ecosystems provide advertising and analytics services that collect, aggregate, process, and trade a rich amount of consumers' personal data and carry out interest-based ad targeting, which raised serious privacy risks and growing trends of users feeling uncomfortable while using the internet services. In this paper, we address users' privacy concerns by developing an optimal dynamic optimisation cost-effective framework for preserving user privacy for profiling, ads-based inferencing, temporal apps usage behavioral patterns, and interest-based ad targeting. A major challenge in solving this dynamic model is the lack of knowledge of time-varying updates during the profiling process. We formulate a mixed-integer optimisation problem and develop an equivalent problem to show that the proposed algorithm does not require knowledge of time-varying updates in user behavior. Following, we develop an online control algorithm to solve the equivalent problem and overcome the difficulty of solving nonlinear programming by decomposing it into various cases and to achieve a trade-off between user privacy, cost, and targeted ads. We carry out extensive experimentations and demonstrate the proposed framework's applicability by implementing its critical components using POC (Proof Of Concept) `System App'. We compare the proposed framework with other privacy-protecting approaches and investigate whether it achieves better privacy and functionality for various performance parameters.\end{abstract}


\titlepgskip=-15pt

\maketitle

\setcounter{page}{38664}

\section{Introduction} \label{section-introduction}

\let\thefootnote\relax\footnotetext{The associate editor coordinating the review of this manuscript and approving it for publication was Junggab Son.}
The online advertising companies enable user tracking with the help of millions of mobile applications (\textit{apps}) offered via various \textit{app} markets, profile users, and enable \textit{targeted advertising} where user's personal information plays an important role. The advertising ecosystems connect billions of mobile devices, including smartphones, tablets, and computer tablets, with the help of various tracking technologies to collect, process and map, and disseminate users' private sensitive information. With the strict laws imposed by the governments, in addition to an abundance of anti-tracking tools/policies, the advertising companies are gradually obsoleting the use of third-party cookies used for (interest-based) ad targeting. Google announcement on Chrome's `Cookie Apocalypse' to phase out support for third-party cookies by 2022\footnote{\url{https://www.adviso.ca/en/blog/tech-en/cookie-apocalypse/}}, similarly, Apple\footnote{\url{https://junction.cj.com/article/button-weighs-in-what-does-apples-idfa-opt-in-overhaul-mean-for-affiliate}} announced to update their tracking by introducing IDFA (Identification for Advertisers) opt-in overhaul, which will have a significant impact over mobile ads targeting and \textit{in-app} data attribution. Hence, instead of relying on second- (i.e. data sharing among selected partners) or third-party data; the advertisers and publishers tend towards first-party data (i.e. owned by the publishers), which has its associated privacy concerns and will limit targeting the vast majority of consumers. The advertisers and publishers are shifting towards maintaining Data Management Platforms (DMPs) and Demand-Side Platforms (DSPs)\footnote{DMP is a unified and centralised technology platform used for collecting, organising, and activating large sets of data from disparate sources. DSP allows for advertisers to buy impressions across several different publisher sites, all targeted to specific users based on key online behaviors and identifiers. See \url{https://www.lotame.com/dmp-vs-dsp/} for detailed discussion over DMP and DSP.} to brand their data and measure performance in a \textit{cookie-less} world. In addition, users are more concerned about their data in the long term e.g. \textbf{D}ata they leave \textbf{a}fter \textbf{d}eath (\textit{Dad}).

An important factor in online targeted advertising is to deliver relevant ads to users to achieve better view/click-through rates without exposing their private and sensitive information, within or outside the advertising ecosystem. In its current working mechanism, user tracking and advertising have raised significant privacy concerns, as suggested in several studies \cite{ullah2020privacysurvey, wang2020aiming, mamais2019privacy, liu2016privacy, rafieian2020targeting, beigi2019protecting}. Other works show the extent to which consumer's activities are tracked by third-parties and across multiple \textit{apps} \cite{razaghpanah2018apps}, mobile devices leaking Personally Identifiable Information (PII) \cite{elsabagh2020firmscope, ren2016recon}, accessing user's sensitive information via APIs \cite{verderame2020reliability}, and profile inference attacks based on monitoring ads \cite{tchen2014}. Research studies indicate, unless consumers have specifically consented to it, that the ad/tracking companies evaluate user behavior and tailor advertising to reach specific audiences. The American self-regulatory authority, AdChoices\footnote{\url{https://optout.aboutads.info/?c=2&lang=EN}} program, presents a platform for consumers to \textit{opt-out} of targeted ads and tracking done by the participating companies in WebChoices. However, this would result in less revenue, in addition to, presenting less relevant ads and lower click-through rates, as evident in \cite{johnson2020consumer}.

This paper contributes in several ways; the proposed framework \textbf{protects user's privacy} in the mobile advertising ecosystem from `profiling of specific private attributes' that are potentially used by the analytics companies for tracking user's behavior (over `web and apps usage behavior') and controls the magnitude of `ads targeting' based on these attributes. To provide profiling privacy, we note that the private attributes (i.e. user's private interests) during the profiling process dominate user profiles, which can be lowered down to reduce its dominating factor i.e., to produce disturbance in user profiles and hence to control `interest-based targeting' and temporal changes in the user profiling process. The proposed framework provides \textbf{an optimal privacy-preserving user profiling that is cost-effective and achieves a trade-off between user privacy and targeted ads}. We recommend the use of (other than the set of installed) \textit{apps} to flatten the usage pattern of users to protect temporal user's \textit{apps} usage behavior, which would be of particular scenario when users use mobile phones during business hours. Ideally, these recommended \textit{apps} can be run when there is no user activity to achieve an average usage of \textit{apps} to actual usage patterns. However, a major challenge is to evaluate temporal changes in user profiles and \textit{apps} usage i.e. respectively, the lack of knowledge of time-varying updates in user profiles, which we classify as updates caused by `browsing history/searches', `interactions with ads e.g. view/click', and the types of `apps installed and used' \textit{and} `user's time-varying future apps usage behavior'.

Subsequently, using Lyapunov optimisation, we \textbf{develop an optimal control algorithm} for identifying updates in user profiles and the user's usage behavior of actual \textit{apps} to capture the temporal changes during the profiling process; note that other optimisation mechanisms for dynamic systems can also be used e.g., \cite{neely2012dynamic}. User profiles consist of various \textit{interests}, gathered in similar categories, that are derived based on the user's (private) behavioral activity from utiilising installed \textit{apps}, activities over the web, and their interactions with ads. Our purpose is to protect attacks on privacy, via \textit{app}-based profiling (i.e. context profiling) and ad-based profile inferencing (i.e. user profiling based on targeted ads), of selected private (that may be considered private by the user) \textit{interest} categories in a user profile. E.g. the user may not wish the categories of gambling or porn to be presented in their profiles or to be targeted with relevant ads, which would of particular relevance when business devices are used for private purposes. In addition to privacy protection, which imposes cost overhead by running the recommended \textit{apps}, the \textbf{proposed framework minimises the cost introduced}, termed as `resource overhead', by bounding the use of new \textit{apps} among various boundaries of weightage assigned to interests categories in user profiles.

Furthermore, we \textbf{investigate the profiling process used by Google AdMob}, a major analytics network, during \textit{establishing} user profile and during when profile \textit{evolves} over time, by investigating the relation between mobile \textit{apps} characteristics and the resulting profiles. We \textbf{carry out extensive profiling experiments} by 1.) examining the contribution of individual \textit{app} in a user profile, 2.) experiments with the recommended \textit{apps} for protecting user privacy for `apps usage behavior' and evaluate their effect over user profiles, and 3.) experiments for evaluation of resource overheads; overall these experiments were run for over 5 months. Our experiments show that the mapping of interest categories can be predicted from installed and used \textit{apps} with an accuracy of 81.4\% along with the private and dominating interests categories based on user's activity over a mobile device. In addition, using these experiments, we found that the profiling interests are selected from a pre-defined set of interests categories and that the \textit{apps} to interests mapping is deterministic, which requires a specific amount of time (up to 72 hours) and a certain level of activity to \textit{establish} a user profile.

We \textbf{propose various changes in `User Environment' in a mobile advertising ecosystem} i.e. we suggest `System App' with the following functionalities: It implements local user profiling based on `user's apps usage behavior', `browsing history/searches' and `user's interactions with ads'; `System Engine' that keeps local repository and determines recommended \textit{apps}; Protects user's privacy for their private sensitive attributes; Implements proposed online control algorithm for jointly optimising user privacy and associated cost. Furthermore, we \textbf{implement a POC (Proof Of Concept) `System App' of our proposed framework} to demonstrate its applicability in an actual environment. As opposed to the legacy advertising operations where users are tracked for their activities; the `System App' passes anonymous \textit{apps} usage info/statistics and generated profiles to the analytics servers within the advertising system. In addition, the analytics server in the current advertising system only evaluates stats for \textit{apps} usage, which are recorded for \textit{billing} both for ad system and \textit{apps} developers.

Finally, we \textbf{provide a hypothetical discussion} over the use of other privacy protection mechanisms e.g., \textit{differential privacy}, \textit{cryptographic approaches} etc., in the mobile advertising environments and compare the proposed framework with these privacy protection mechanisms for various performance parameters.

The paper is organised as follows: Related work is presented in Section \ref{related-work}. Section \ref{problem-formulations} presents background on user profiling and the ads ecosystem, proposed addressed problem, and threat model. In Section \ref{system-models}, we present a system model and investigate the profile creation and user profiling process. Section \ref{optimal-profiling} presents an optimal privacy-preserving system and further discusses the proposed framework. The optimal control algorithm is discussed in Section \ref{online-control-algo}. Various evaluation metrics are discussed in Section \ref{perfor-analysis}. Section \ref{evaluation} discusses system evaluation, our experimental setup and results. We further discuss the applicability of the proposed framework and its comparison with other privacy protection mechanisms in Section \ref{discussion}. Finally we conclude in Section \ref{conclusion}.

\section{Related Work} \label{related-work}
In our recent work \cite{ullah2020privacysurvey} we provide a detailed survey on privacy leakages in the profiling process, leakage of personal information by advertising platforms and ad/analytics networks, the measurement analysis of \textit{targeted advertising} based on user's interests and profiling context, compare various privacy-preserving advertising systems for various capabilities, such as the underlying architecture used, the privacy mechanisms and the deployment scenarios, furthermore, we present detailed discussion over ads delivery process in both \textit{in-app} and \textit{in-browser targeted ads}. Privacy threats in \textit{targeted advertising} have been extensively studied in literature e.g., direct and indirect (inferred) leakage of private information \cite{mamais2019privacy, englehardt2016online, frik2020impact, meng2016price} or third-party ad tracking and visiting \cite{shuba2020nomoats, iqbal2020adgraph, merzdovnik2017block, das2018web}. These works show the prevalence of user (from their online data) tracking on both web and mobile environments and demonstrate the possibility of inferring user's PII, such as email addresses, age, gender, relationship status, etc. The ad/analytics libraries (embedded with mobile \textit{apps}) leak users' personal information to the ad system and (is more likely to) third-party tracking, which is systematically collected by such libraries. Such example works \cite{liu2019privacy, taylor2017intra} study information collected by analytics libraries integrated within mobile \textit{apps} and leaking of private information. Similarly, the authors in \cite{grace2012unsafe} show privacy and security risks by analysing 100,000 Android \textit{apps} and find that majority of the embedded libraries collect and share private information. Other studies \cite{book2013case, demetriou2016free} find that majority of the mobile \textit{apps} do not implement private APIs and send private information to ad servers.

Recall that user profiling and \textit{ads targeting} is carried out based on user's activity over \textit{apps} and internet; the authors in \cite{book2015empirical} manually create user-profiles and examine that majority of \textit{targeted ads} were based on the fabricated profiles. In our previous work \cite{ullah2014characterising}, and later supported by another study \cite{nath2015madscope}, we examine various tracking information sent to the ad networks and the level of \textit{ads targeting} based on such profiling information. Another work \cite{meng2016price} investigates the information collected by ad networks using installed \textit{apps} and reverse engineer the \textit{targeted ads}. In line with the above works, we also study \cite{tchen2014} leakage of private sensitive information by examining actual network traffic through vulnerabilities in mobile analytics services, reconstructing the exact user profile of several participating volunteers, and further studying its influence over \textit{ads targeting}. This study was mainly for Google Mobile Analytics and Flurry Analytics. Third-party tracking also actively collect, manage and distribute user's sensitive information, which has been widely studied in literature e.g. \cite{binns2018third, lerner2016internet, binns2018measuring, pan2015not, roesner2012detecting, merzdovnik2017block} study the distribution of third-party trackers across the web and Android \textit{apps} and their impact on user privacy.

Other works \cite{ullah2020privacy, ullah2014profileguard, haddadi2010mobiad, backes2012obliviad, Hardt:2012}, suggest the \textit{app}-based user \textit{profiling}, stored locally on mobile device. There are various \textit{in-app} privacy-preserving mobile advertising and personalisation proposals, such as, Adnostic \cite{toubiana2010adnostic}, Privad \cite{guha2011privad}, RePriv \cite{fredrikson2011repriv}, MobiAd \cite{haddadi2010mobiad}, Splitx \cite{Chen:2013:Splitx}, ProfileGuard \cite{ullah2014profileguard}. The ProfileGuard is an \textit{app}-based profile obfuscation mechanism for protecting user privacy using various obfuscation strategies. \hl{However, this work only protects user privacy by reducing dominance level of particular profiling interests, which has greater impact over the targeted ads and possibly would attract majority of the irrelevant ads.} Similarly, there are some other solutions that are crypto-based implementation of various techniques under Private Information Retrieval (PIR) \cite{ullah2020privacysurvey} and Blockchain-based solutions \cite{ullah2020privacy, gu2018secure} for decentralised advertising system that enables private profiling and \textit{targeted ads}.

\hl{We note that majority of the \textit{app}-based privacy protection mechanisms protect user privacy based on profiling interests whereas disregarding various other factors; such as user privacy based on an application profile, the types of \textit{apps} installed and used along with their time-varying future usage behaviour, frequency of \textit{apps} usage, users' web searches, and user interactions around the ads e.g., clicks. Furthermore, we note that these works do not consider the trade-off between user privacy, the targeted ads, and the cost to achieve user privacy.} This work jointly optimises \textit{user's profiling privacy}, the \textit{user's apps usage behavior}, and \textit{cost} of achieving the level of privacy. We provide an online control algorithm that provides a trade-off by achieving between user \textit{privacy} and \textit{targeted ads}. This was achieved by optimally identifying updates in user profiles and user's \textit{apps} usage behavior of actual \textit{apps} by capturing temporal changes during the profiling process. We mainly address the attacks regarding \textit{app}-based profiling, ads-based inferencing, and analysing user's behavior by sniffing network traffic of legitimate users.

\section{Problem Formulation} \label{problem-formulations}

The Advertising and Analytic (A\&A) companies rely on users' tracking data to profile and to \textit{target} them with relevant advertisements, both `Web \& App'-based to cover the vast majority of the audience of diverse interests, to increase their advertising market revenue. This exposes sensitive information about users, in particular, their online behavior e.g., web browsing/searching, or when profiling is based on \textit{apps} representing sensitive and damaging information, e.g., gambling problems indicated by a game \textit{app}, or the mobile \textit{apps} usage behaviors e.g. playing games or use of gambling \textit{apps} early morning in bed or during lunch break in-office hours. We then present the privacy issues related to the profiling process for both `Web \& App' activity and \textit{app} usage behavior, subsequently, we present the problem and threat model discussed in this paper.

\subsection{User profiling}\label{user-profiling}
The advertising companies, e.g., Google, profile users based on information a user adds to the Google account, its estimation of user's interests based on the mobile \textit{apps} usage and web histories, and data collected from the partnering analytics companies with Google, which effectively carry out \textit{targeted adverting} based on the personal information extracted through various tracking technologies.

\begin{figure}[h]
\begin{center}
\includegraphics[scale=0.25]{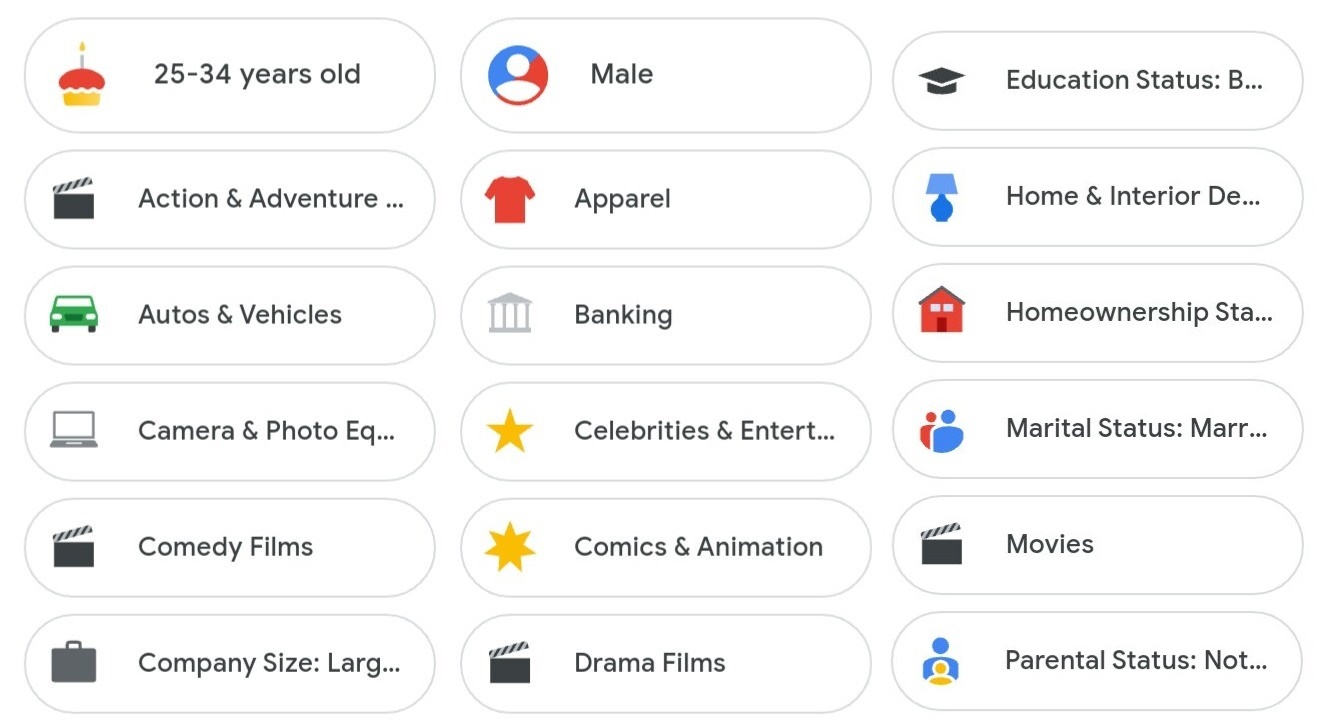}
\caption{An (anonymous) example user profile estimated by Google as a results of `Web \& App' activity.}
\label{example-profile}
\end{center}
\end{figure}

Figure \ref{example-profile} shows an example user profile\footnote{Google AdMob profile can be accessed through the \textit{Google Settings} system \textit{app} on Android-based devices; accessible through \texttt{Google Settings} $\to$ \texttt{Ads} $\to$ \texttt{Ads by Google} $\to$ \texttt{Ads Settings}.}, estimated by Google\footnote{Profiling interests can be found here: \url{https://adssettings.google.com/authenticated}.}, which consists of demographics (e.g. gender, age-ranks) and profiling interests e.g. Action \& Adventure Films. We call this an \textit{Interests profile} with all those interests defined by A\&A companies (e.g. Google) and is used by them to individually characterise user's interests across the advertising ecosystem. Similarly, we introduce \textit{Context profile} that is the combination of \textit{apps} installed from various categories, e.g. Games, Entertainment, Sports etc., on a user mobile device; detailed discussion over \textit{Context profile} is given in Section \ref{system-models}. We note, using our extensive experimentations \cite{ullah2014profileguard}, that \textit{Context profile} profile is also (partly) used by the analytics companies to individually characterise user's interests across the advertising ecosystem.

Furthermore, the \textit{ads targeting} is based on demographics to reach a specific set of potential customers that are likely to be within a specific age range, gender, etc. Google\footnote{Demographic Targeting: \url{https://support.google.com/google-ads/answer/2580383?hl=en}} presents a detailed set of various demographic \textit{targeting} options. The demographics are usually grouped into different categories, with specific options such as age-ranges, e.g. `18-24', `25-34' etc., and gender e.g., `Male', `Female', and other demographic \textit{targeting} options e.g. parental status, location etc. We note that this profiling is the result of interactions of user devices with AdMob \texttt{SDK} \cite{ullah2014profileguard} that communicates with Google Analytics for deriving user profiles. A complete set of `Web \& App' activities of an individual user can be found under `My Google Activity'\footnote{\url{https://myactivity.google.com/myactivity?otzr=1}}, which helps Google make services more useful.

\begin{figure*}[h]
\begin{center}
\includegraphics[scale=0.6]{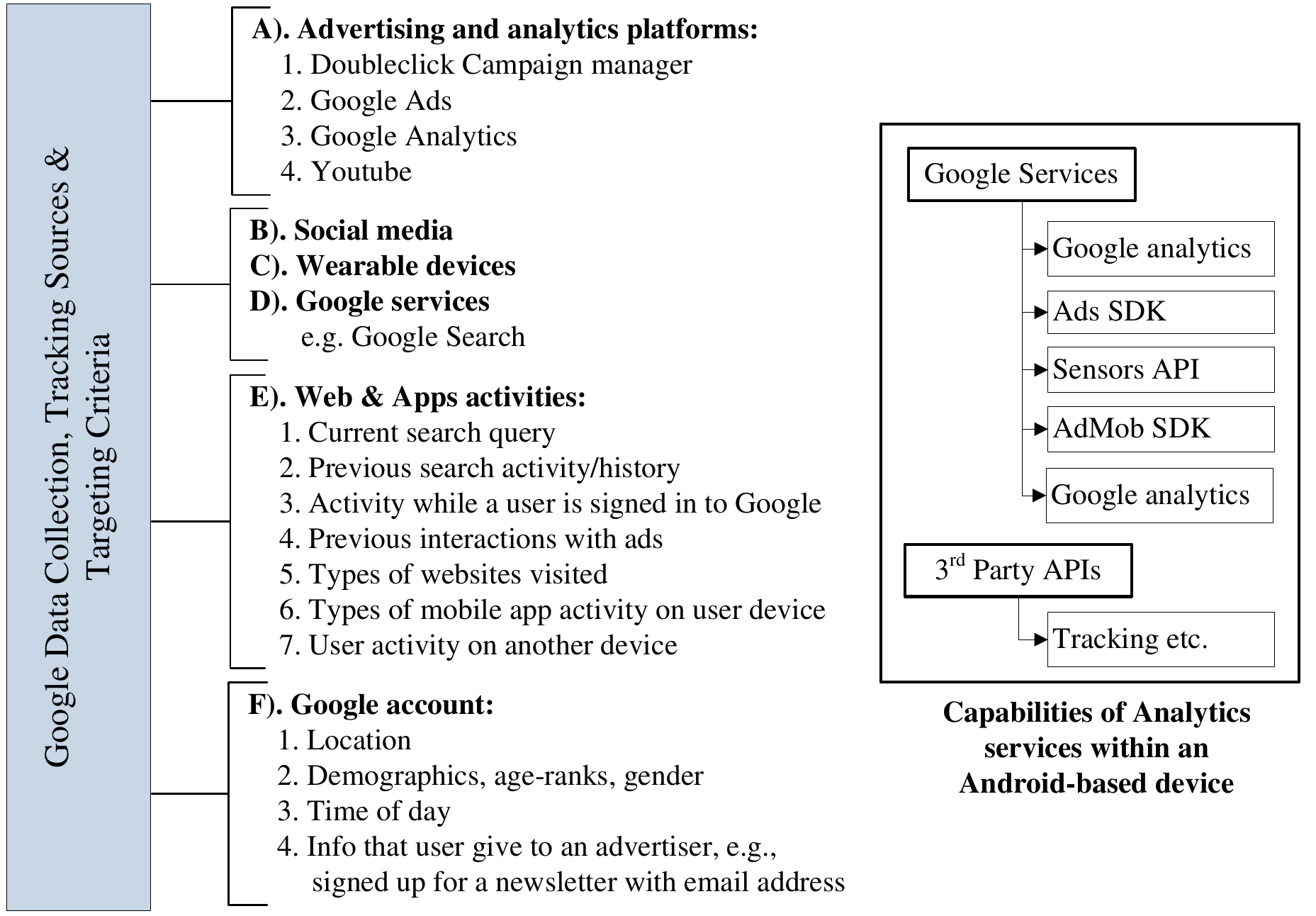}
\caption{Google's data collection and tracking sources for targeting users with personalised ads (left) and tracking capabilities of analytics libraries enabled within mobile devices (right).}
\label{data-extraction}
\end{center}
\end{figure*}

Figure \ref{data-extraction} shows various sources/platforms that Google uses to collect data and \textit{target} users with personalised ads. These include a wide range of different sources enabled with tracking tools and technologies, e.g., the `Web \& App' activities are extracted with the help of \texttt{Andoird/iOS SDK}s, their interactions with Analytics servers within the Google network, cookies, conversation tracking\footnote{\url{https://support.google.com/google-ads/answer/6308}}, web searches, user's interactions with presented ads, etc. Similarly, Google's connected home devices and services\footnote{Google's Connected Home Devices and Services: \url{https://support.google.com/googlenest/answer/9327662?p=connected-devices&visit_id=637357664880642401-2675773861&rd=1}} rely on data collected using cameras, microphones, and other sensors to provide helpful features and services\footnote{Sensors in Google Nest devices: \url{https://support.google.com/googlenest/answer/9330256?hl=en}}. The tracking data (up to several GBs of data), personalised for individual users, can be exported using Google \texttt{Takeout}\footnote{Your account, your data: \url{https://takeout.google.com/}} for backup or use it with a service outside of Google. This includes the data from a range of Google products, such as email conversations (including Spam and Trash emails), contacts, calendars, browsing and location history, and photos.

\subsection{Entities within advertising system}\label{problem-statement}
\begin{figure}[h]
\begin{center}
\includegraphics[scale=0.45]{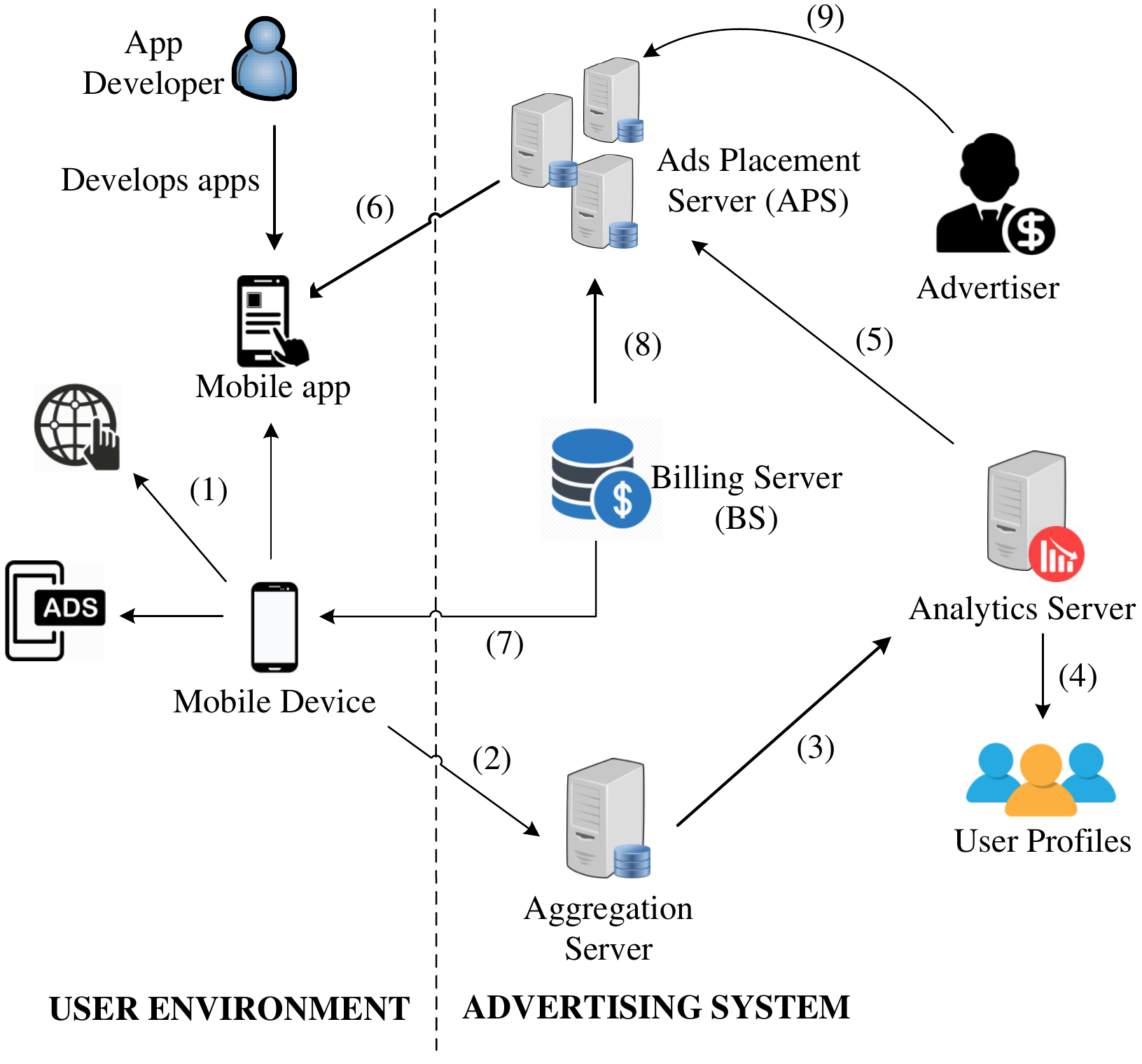}
\caption{The advertising system consists of user and advertising environments. Following functionalities: (1) Data collection and tracking, (2) Send tracking data to \texttt{Aggregation} server, (3) Forward usage info to \texttt{Analytics} server, (4) User profiling, (5) Send profiling info to \texttt{APS}, (6) Deliver targeted/generic ads, (7) Billing for Apps Developer, (8) Billing for Ad System, (9) Advertiser uploads ads, who wishes to advertise with Ad system.}
\label{basic-system-model}
\end{center}
\end{figure}

Figure \ref{basic-system-model} shows a representative \textit{in-app} mobile advertising ecosystem showing the information flow among its different parties for profiling users for enabling \textit{targeted ads} based on `Web \& App' activities. User usually install a number of \textit{apps} on their mobile devices, that are utilised with specific frequency. The mobile \textit{apps} include analytics \texttt{SDK}, which directly reports user's activity (as mentioned in above section) and sends ad requests to the analytics and ad network. Various advertising entities play important role in enabling tracking and ads dissimilation in an ad system, comprises the \texttt{Aggregation}, \texttt{Analytics}, \texttt{Billing}, and the \texttt{Ads Placement} servers. Collected tracking data is used by the \texttt{Analytics} server that constructs \textit{Interests profiles} (associated with specific mobile devices and corresponding users) with specific profiling interests related to user's (private) behavior. The \textit{targeted ads} are served to mobile users according to their (individual) profiles. We note that other i.e., \textit{generic} ads are also served \cite{ullah2014characterising}. The \textit{Billing} server includes the functionality related to monetising \textit{Ad impressions} (i.e. ads displayed to the users in specific \textit{apps}) and \textit{Ad clicks} (user action on presented ads).

\subsection{\hl{Research  Statement}}\label{problem-statement}
The problem addressed in this paper is where the A\&A companies track mobile users for their activities, profile them (inferred via relationships among individuals, their monitored responses to previous advertising activity and temporal behavior over the Internet), and \textit{target} them with ads specific to individual's interests. The user profiling and \textit{ads targeting} expose sensitive information about users \cite{ullah2020privacysurvey}, e.g. the \textit{target} could browse through medical related websites or \textit{apps}, revealing (including a third-party, such as the website owner) to the advertising systems that the user has medical issues.

\begin{figure}[h]
\begin{center}
\includegraphics[scale=0.65]{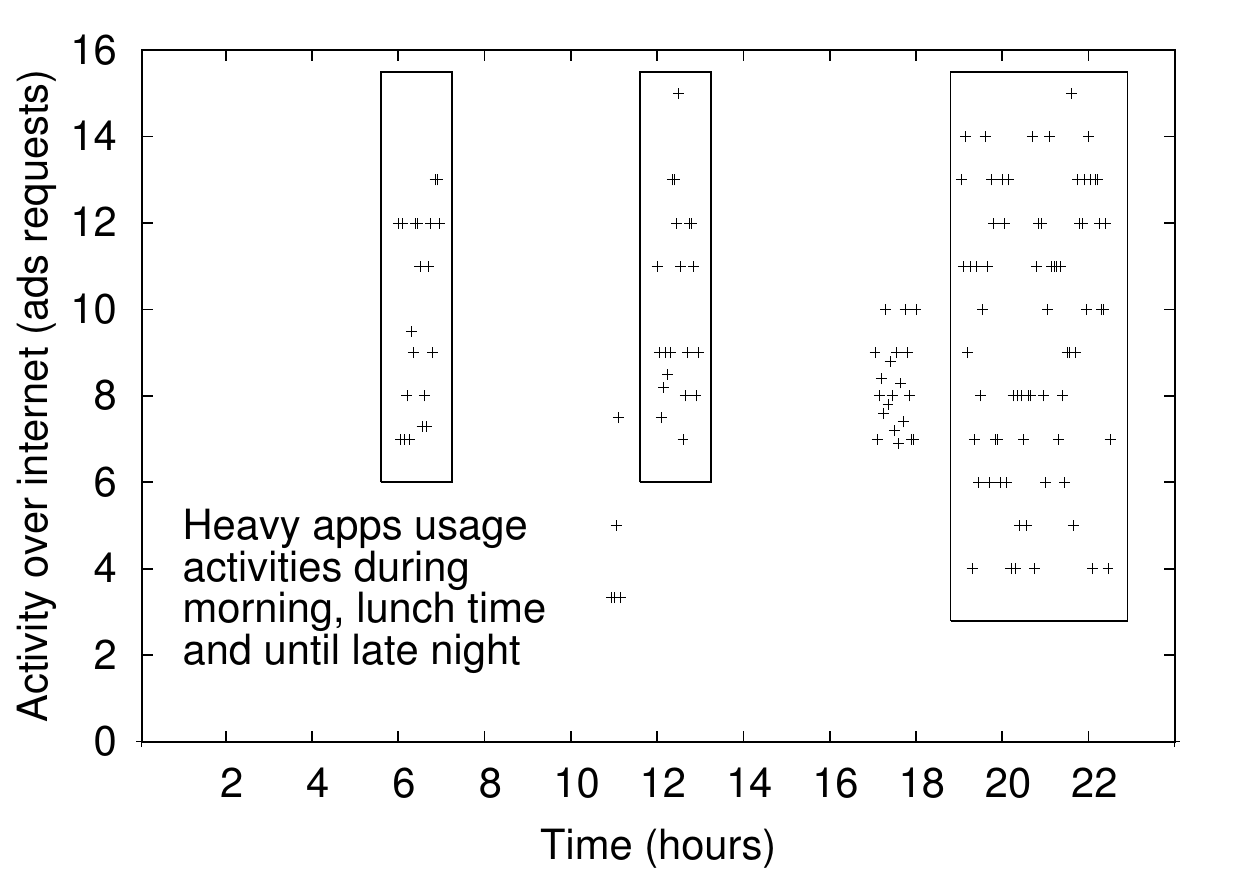}
\caption{Apps usage profile for typical day showing \textit{apps} usage activity during 24 hours.}
\label{apps-usage-behavior}
\end{center}
\end{figure}

Furthermore, we address the privacy issues where an adversary (either the analytics companies examining user activity or an intruder listening to the ad or control traffic) can determine the \textit{app}'s usage activity e.g., someone plays games during late night or early morning and their activity is intercepted by their neighbors. Note that the user's \textit{apps} usage activities can be exposed by intercepting the \textit{apps} communication using the connected network \cite{tchen2014}, in addition, users' activities are exposed to the advertising systems during their interactions using ad/analytics \texttt{SDK}s. Figure \ref{apps-usage-behavior} shows an example \textit{apps} usage profile (in plaintext) for a typical day, showing a direct threat to user's privacy for \textit{apps} usage.

In particular, we address three privacy attacks: \textbf{1.} \textit{Legitimate user profiling} by A\&A, the user profiling is implemented via an analytics \texttt{SDK} for reporting user's activities to A\&A companies, hence, intercepts various requests to/from mobile users. \textbf{2.} \textit{Indirect privacy attack}, involves third parties that could intercept and infer user profiles based on \textit{targeted ads}. \hl{We note that an adversary (e.g., an intruder listening to ad traffic) can determine the \textit{app}'s usage activities (e.g., user using gambling \textit{apps}), which can be exposed by intercepting the \textit{apps} communication of the connected network} \cite{tchen2014}. \hl{In addition, the adversary may intercept users' other interactions e.g., interactions with ads for views/clicks, web searches, data communicated via connected devices or sensors.} \textbf{3.} \textit{Apps usage behavioral attack} to know user's \textit{apps} usage activities. \hl{Alternatively, such \textit{apps} usage activities are exposed to the A\&A companies during their interactions using ad/analytic \texttt{SDK}s, embedded in mobile apps} \cite{ullah2014characterising, ullah2020privacysurvey}. \hl{For this reason, using \textit{Lyapunov} optimisation, we develop an optimal control algorithm for identifying updates in user profiles based on \textit{apps} usage behavior for their temporal changes during the profiling process.} We presume that the users do not want to expose their private interests to adversaries (including advertising agencies) and are willing to receive relevant ads based on their interests.

\subsection{Threat Model}\label{threat-model}
Our primary goal in this work is to achieve an \textit{optimal privacy-preserving profile management} that preserves \textit{user's privacy due to profiling interests derived via apps usage}, privacy in terms of \textit{apps usage behavior}, \textit{users' web history/searches}, and their \textit{interactions with ads}. We start by analysing and developing the profiling process, identify the dominating interests that expose user privacy and could affect \textit{ads targeting}, the proportion of interests in user profile, and the \textit{apps} usage activity. We further compare proposed solution and its applicability with other privacy preserving mechanisms such as \textit{differential privacy}, \textit{anonymisation}, \textit{randomisation}, \textit{profile-based obfuscation} \textit{Blockchain-based solutions} and \textit{crypto-based} mechanisms e.g., \textit{private information retrieval}) in an advertising scenario for the problems addressed in this work. Finally, we evaluate the trade-off between the achieved user privacy, cost of achieving privacy and \textit{targeted ads}. Hence, in this paper, we jointly optimise the user's privacy due to \textit{profiling interests}, the \textit{cost of achieving privacy}, and the \textit{apps usage behavior}.

\section{System Components} \label{system-models}

We formalise the system model that consists of the \textit{apps}' profiles, interests' profiles, and the conversion of resulting profiles by use of applications in an \textit{app} profile. In particular, we provide the insights of \textit{establishment} of \textit{Interests profiles} by individual \textit{apps} in the \textit{Context profiles} and then show how the profiles \textit{evolve} when some \textit{apps} other than the initial set of \textit{apps} are utilised. 

\subsection{System Model}\label{system-model}
We denote an \textit{app} by ${a_{i,j}}$, $i=1,...,A_j$, where $A_j$ is the number of \textit{apps} that belong to an \textit{app} category $\Phi _j$, $j=1,..., \phi $ and $\phi $ is the number of different categories in a marketplace (e.g., in Google Play or in the Apple App store). \hl{For example, there are several \textit{apps} categories $\phi$ in Google Play Store\footnote{\url{https://play.google.com/store}}, such as `Art \& Design', `Books \& Reference', `Entertainment' etc.; we reference each category with an index $j$. Similarly, an individual category $j$ may have numerous applications, which can be downloaded and used, e.g., the `LinkedIn Learning: Online Courses to Learn Skills' \textit{app} is categorised under the category `Education', which is represented with the index $i$. Hence, for the \textit{app} `LinkedIn Learning: Online Courses to Learn Skills', the ${a_{i,j}}$ can be interpreted as `this \textit{app} is indexed $i$ and is categorised under the $j^{th}$ category'. }

In addition, we note that an individual \textit{app} ${a_{i,j}}$ is characterised by a set of keywords \({\kappa _{i,j}}\) that includes \textit{app}'s category (e.g. Business, Entertainment etc.) and dictionary (specific to \textit{this} particular \textit{app}) keywords from its description. We represent with $\mathcal{A}$ the entire set of mobile \textit{apps} in any marketplace, organised in various categories.

A user may be characterised by a combination of \textit{apps} installed on their mobile device(s), comprising a subset $S_a \in \mathcal{A}$. \hl{For example, a subset $S_a$ may comprise of various \textit{apps} e.g., `LinkedIn', `Outlook', `Uber', `Zoom' etc.} Subsequently, the \textit{Context profile} ${K_a}$ can be defined as:

\begin{equation}\label{initial-app-profile}
{K_a} = \left\{ {\left\{ {{a_{i,j}},{\Phi _j}} \right\}:{a_{i,j}} \in {S_a}} \right\}
\end{equation}

The A\&A companies, such as Google or Flurry, partly profile and \textit{target} users based on the combination of mobile \textit{apps} installed on their devices i.e. \textit{Context profile}. We have the following constraints on the user's \textit{Context profile}:

\begin{eqnarray}\label{eq:apps-constraints}
0 < n\left( {{K_a}} \right) \le \mathcal{A} \nonumber \\
n\left( {{K_a}} \right) \ge 1 \\
{\Phi _j}\left( {{a_{i,j}}} \right) \ne 0 \nonumber \\
\phi \subseteq {\Phi _j}\,\,;\,\,\forall j \nonumber 
\end{eqnarray}

The $n\left( {{K_a}} \right)$ is the total number of \textit{apps} installed on a device, hence, using Eq. (\ref{eq:apps-constraints}), it is important to make sure that the $n\left( {{K_a}} \right)$ should not exceed the total number of available \textit{apps}; that a mobile device must have at least one \textit{app} installed for \textit{contextual targeting}; that an \textit{app} $a_{i,j}$ must belong to any of the specified categories; and that the \textit{app}'s category is not undefined within an \textit{app} market. Various important notations and their descriptions are presented in Table 1.

\subsection{Representing apps in user profile}\label{representing-user-profile}
We note that the A\&A companies classify users by defining a set of profiling interests $\mathcal{G}$, i.e., characteristics that may be assigned to users i.e. an \textit{Interest profile}. E.g. Google profile interests\footnote{Google profile interests are listed in \url{https://adssettings.google.com/authenticated?hl=en}, managed under the 'How your ads are personalized'; other options and Google services can also be verified on Google Dashboard \url{https://myaccount.google.com/dashboard?hl=en}.} are grouped, hierarchically, under various interests categories, with specific interests. We denote, using ${g_{k,l}}$, $k=1,...,G_l$, where $G_l$ is total number of interests that belong to an interest category $\Psi_l$, $l=1,..., \psi $. $\psi $ is a total number of interest categories defined by analytics companies. An individual interest ${g_{k,l}}$ consists of a set of keywords \({\kappa _{k,l}}\), which characterises specific interests. Following, to enable \textit{interest targeting}, represents an \textit{Interests profile} ${I_g}$, consists of a subset $S_g$ of specific interests:

\begin{equation}\label{eq:initial-interes-profile}
{I_g} = \left\{ {\left\{ {{g_{k,l}},{\Psi _l}} \right\}:{g_{k,l}} \in {S_g}} \right\}
\end{equation}

Although various types of information may be used to generate user profiles, as shown in Figure \ref{data-extraction}, however, our focus is mainly on \textit{installed and used/unused apps}\footnote{Full list of installed \textit{apps} is located at: \url{/data/data/com.android.vending/databases/library.db}.}, the \textit{history resulted from web browsing or web searches}, and the \textit{clicked ads} i.e. collectively described as the `Web \& Apps' activity. Similarly, other \textit{targeting criteria} can also be represented in an \textit{Interest profile} as a specific interest e.g. \textit{demographics}, \textit{contributing analytic platforms}, \textit{social media}, and other \textit{Google services}. An example \textit{demographics interests} is also shown in Figure \ref{example-profile}.

\begin{figure}[h]
\begin{center}
\includegraphics[scale=0.45]{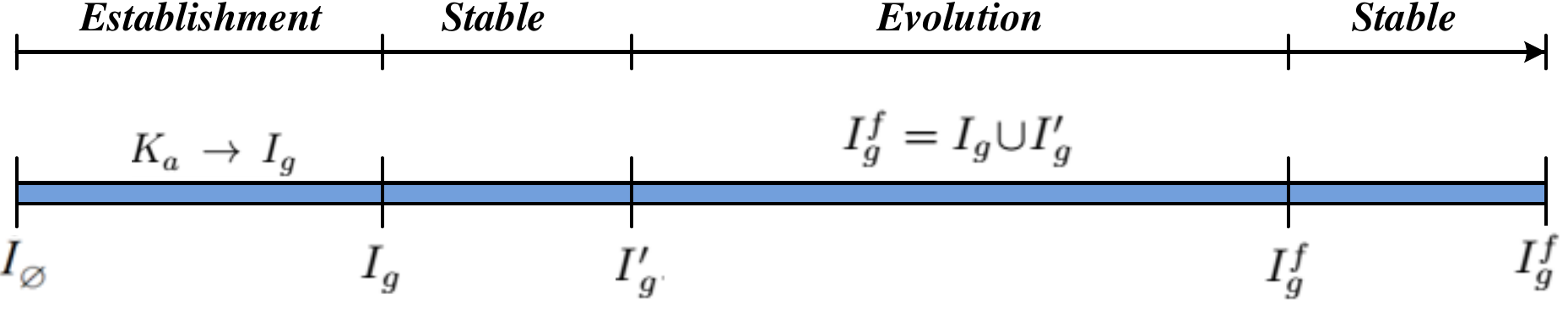}
\caption{Profile \textit{establishment} \& \textit{evolution} processes. ${I_\emptyset }$ is the empty profile before any activity over user device. During the \textit{stable} states, the \textit{Interest profiles} $I_g$ or $I_g^f$ remains the same and further activities of the same \textit{apps} have no effect over the profiles \cite{ullah2020privacy}.}
\label{profile-development}
\end{center}
\end{figure}

The \textit{targeting} components are eventually grouped in \textit{Interest profile} that are used for \textit{ads targeting}. The \textit{Interest profiles} undergo different processes e.g. the \textit{profile establishment} process (i.e. the derivation of \textit{Interests profile} by \textit{apps}; \({K_a} \to {I_g}\)) (generating a specific set of interests \({S_{{K_a}}}\left\{ {{\Psi _l}} \right\}_{l = 1}^\varepsilon \) using a \textit{mapping} function $f$ i.e. \(f:{\Phi _j} \to {\Psi _l} = {S_{{K_a}}}\left\{ {{\Psi _l}} \right\}_{l = 1}^\psi \)), the \textit{profile evolution} $ I{'_g}$ process (i.e. each time variations in the users' behaviour are observed), and \textit{profile development} process (i.e. the minimum level of activity required to develop a \textit{stable} user profile $I_g^f$) i.e. $I_g^f = {I_g} \cup I{'_g}$. Figure \ref{profile-development} shows user profiling process, detailed discussion over these various processes of user profiling can be found in \cite{ullah2020privacy}. Furthermore, detailed experimental evaluations over insights on profiling rules, \textit{apps} usage threshold during \textit{profiling establishment}, and \textit{mapping} rules for \textit{Context profile} to \textit{Interest profiles} can be found in \cite{ullah2014profileguard}.

\begin{table*}[t]
\begin{center} 
\label{table:notations}
{\begin{tabular}{c|l}
\hline
 \textbf{Symbols}& \textbf{Description}\tabularnewline
\hline
${\mathcal{A}}$& Available \textit{apps} in a marketplace \tabularnewline
\hline
$\phi$ & Total number of \textit{apps} categories, $\Phi_j$ is a selected category, $j=1,...,\phi$ \tabularnewline
\hline
 ${S_a}$& Subset of \textit{apps} installed on a user's mobile device\tabularnewline
\hline
 ${a_{i, j}}$& An \textit{app} ${{a_{i,j}} \in A}$, $i=1,...,A_j$, $j=1,...,\phi$, $A_j$ is the number of \textit{apps} in $\Phi_j$\tabularnewline
\hline
\({\kappa _{i,j}}\)& Set of keywords associated with individual \textit{app} ${a_{i,j}}$ including its category\tabularnewline
\hline
 $K_a$& App profile consisting of \textit{apps} $a_{i,j}$ and their categories $\Phi_j$ \tabularnewline
\hline
 $\mathcal{G}$& Set of interests in Google interests list\tabularnewline
\hline
 $\Psi_l$& Interest category in $\mathcal{G}$, $l=1,...,\psi$, $\psi$ is the number of interest categories defined by Google\tabularnewline
\hline
 $S_g$& Subset of Google interests in $\mathcal{G}$ derived by $S_a$\tabularnewline
\hline
 $I_g$& Interest profile consisting of ${g_{k,l}}$, ${g_{k,l}} \in S_g$\tabularnewline
\hline
 ${g_{k,l}}$& An interest in $I_g$, $k \in G_l$, $l \in \psi$ \tabularnewline
\hline
\({\kappa _{k,l}}\)& Set of keywords associated with individual interest ${g_{k,l}}$ including interest category\tabularnewline
\hline
 $S_g^{i,j}$& Set of interests derived by an \textit{app} $a_{i,j}$ \tabularnewline
\hline
$f$ & Mapping that returns the derivation of \textit{apps} (or history, ad's category) category to interest category \tabularnewline
\hline
\({\eta _l}\left( {{\Psi _l}} \right)\)& The \textit{weightage} assigned to interest category in $\Psi _l$ in a profile with, respectively with \(\eta _l^{\min }\) and \(\eta _l^{\min }\) thresholds \tabularnewline
\hline
\({{\Pi _m}}\)& Profiling components for representing \textit{browsing history/searches} \tabularnewline
\hline
\({{\Gamma _n}}\)& Profiling components for representing \textit{interactions with ads} \tabularnewline
\hline
$t$& Time slot \tabularnewline
\hline
\({C_t}\left( {{\eta _l}\left( {{\Psi _l}} \right)} \right)\)& Tracking change in interest category due to \textit{changes in apps} usage\tabularnewline
\hline
\({C_t}\left( {{\eta _m}\left( {{\Pi _m}} \right)} \right)\)& Tracking \textit{change in web browsing history/searches}\tabularnewline
\hline
\({C_t}\left( {{\eta _n}\left( {{\Gamma _n}} \right)} \right)\)& Tracking \textit{interactions with ads}\tabularnewline
\hline
\({U{'^\tau }\left( {{K_a}} \right)}\) & Usage of \textit{apps} in $K_a$ at time slot $t$; average usage of each \textit{app} is given by \({\overline {U{'^\tau }\left( {{K_a}} \right)} }\)\tabularnewline
\hline
\({\eta _{l'}}\left( {{\Psi _{l'}}} \right)\) & The \textit{weightage} assigned to interest category generated by recommended app(s), where \(l' \ne l\) \tabularnewline
\hline
$S_o$& Set of recommended \textit{apps}\tabularnewline
\hline
\({C^t}\)& Reduction of overall user's time available by use of recommended \textit{apps} to the original \textit{apps}\tabularnewline
\hline
\multirow{2}{*}{\({R^t}\)}& Resource usage of recommended \textit{apps}, in particular, \(R_b^t\), \(R_c^t\), \(R_p^t\) respectively represent the \textit{battery consumption}, \tabularnewline
&\textit{communication}, and \textit{processing} resource usage\tabularnewline
\hline
$ \beta $ & Adjustable parameter to achieve trade-off between user \textit{privacy} and \textit{targeted ads}\tabularnewline
\hline
\({R_l^t}\) & An advertising request reported at $t$ via ad/analytics SDK e.g. an ad request for display/ad click\tabularnewline
\hline
$\varepsilon$ & Adjustable parameter to achieve \textit{apps} usage to preserve privacy of user's \textit{app} usage behavior\tabularnewline
\hline
\(p\left( t \right)\) & The penalty for minimising the upper bound on $C^t$, profiling privacy, and \textit{apps} usage behavior\tabularnewline
\hline
\end{tabular}} 
 \end{center} \caption{List of Notations}
\end{table*}

In order to represent part (along with the dominating interest categories) of each interest category, we assign \textit{weightage} to each category $\Psi_l$ present in the final user profile $I_g^f$; represented as \({\eta _l}\left( {{\Psi _l}} \right)\), under the following two constraints:

\begin{eqnarray}
\eta _l^{\min } \le {\eta _l}\left( {{\Psi _l}} \right) \le \eta _l^{\max }\,\,\,\,\,\,\,\,\,\,\,\,\forall l \in {\Psi _l}|l \in \psi \label{eq:threshold-limits-max}\\
0 < {\eta _l}\left( {{\Psi _l}} \right) \le \eta _l^{\max } \label{eq:threshold-limits}
\end{eqnarray}

Hence, \textit{weightage} given to an interest category \({\eta _l}\left( {{\Psi _l}} \right)\) is within the \(\eta _l^{\min }\) and \(\eta _l^{\max }\) threshold \textit{weightages}. Subsequently, the user profile is represented as, \({I_g} = {\eta _l}\left( {{\Psi _l}} \right),\forall l \in \psi \).

We note, using our extensive \textit{profiling} experimentations \cite{ullah2014profileguard}, that the set of installed \textit{apps}, on a mobile device, do not necessarily contribute during profiling, in which case specific \textit{apps} do not draw any interests i.e. an empty set. In addition, Eq. (\ref{eq:threshold-limits}) ensures that the assigned threshold should be non-negative and cannot exceed the maximum threshold within a user profile.

\paragraph*{Example -- Assigning Weightages} Let a \textit{user} has five mobile \textit{apps} installed on her device i.e. three from category \textit{a} and one from category \textit{b} and \textit{c} each, respectively with the following percentage of usage time\footnote{Use \texttt{adb logcat -v threadtime} on an Android-based device to find the running time of an \textit{app}; \texttt{threadtime} (default) is used to display the date, invocation time, tag, priority, TID, and PID of a thread issuing the message.}: \textit{a = 20\%}, \textit{b = 70\%}, \textit{c = 10\%}; for simplicity we do not consider \textit{apps} that are \textit{installed but not used}, which can be assigned the lowest \textit{weightages} e.g. \(1/n\left( {{K_a}} \right)\) including system installed \textit{apps}. The weightage for each category \({\eta _l}\left( {{\Psi _l}} \right)\) can be evaluated as follows: \({\eta _l}\left( a \right) = {\raise0.7ex\hbox{$3$} \!\mathord{\left/
 {\vphantom {3 5}}\right.\kern-\nulldelimiterspace}
\!\lower0.7ex\hbox{$5$}} + {\raise0.7ex\hbox{${20}$} \!\mathord{\left/
 {\vphantom {{20} {100}}}\right.\kern-\nulldelimiterspace}
\!\lower0.7ex\hbox{${100}$}} = 0.8\); \,\, \({\eta _l}\left( b \right) = {\raise0.7ex\hbox{$1$} \!\mathord{\left/
 {\vphantom {1 5}}\right.\kern-\nulldelimiterspace}
\!\lower0.7ex\hbox{$5$}} + {\raise0.7ex\hbox{${70}$} \!\mathord{\left/
 {\vphantom {{70} {100}}}\right.\kern-\nulldelimiterspace}
\!\lower0.7ex\hbox{${100}$}} = 0.9\); \,\, \({\eta _l}\left( c \right) = {\raise0.7ex\hbox{$1$} \!\mathord{\left/
 {\vphantom {1 5}}\right.\kern-\nulldelimiterspace}
\!\lower0.7ex\hbox{$5$}} + {\raise0.7ex\hbox{${10}$} \!\mathord{\left/
 {\vphantom {{10} {100}}}\right.\kern-\nulldelimiterspace}
\!\lower0.7ex\hbox{${100}$}} = 0.3\). Hence, it requires that the user is supposed to be \textit{targeted} with the ads related to category \textit{b} with highest proportion, followed by \textit{a} and \textit{c}. Note that for simplicity, these \textit{weightages} can be normalised within the range $0-1$; subsequently, the proportion for delivery of \textit{targeted ads} would be: $a=0.40$, $b=0.45$, and $c=0.15$.

\subsection{Representing browsing history/searches and ad-interactions in a user profile}
We respectively represent \({\Pi _m}\) and \({\Gamma _n}\) profiling components as user \textit{history/search} and \textit{ad-interactions} (e.g. \textit{ad click}). Subsequently, we assign \textit{weightages} to both these components:

\begin{eqnarray}
\left( {\eta _l^{\min },0} \right) < {\eta _m}\left( {{\Pi _m}} \right) \le \eta _l^{\max }\,\,\,\,\forall m,\forall l \in \left[ {1,\psi } \right] \label{eq:history} \\
\left( {\eta _l^{\min },0} \right) < {\eta _n}\left( {{\Gamma _n}} \right) \le \eta _l^{\max }\,\,\,\,\forall n,\forall l \in \left[ {1,\psi } \right] \label{eq:ad-interaction}
\end{eqnarray}

Note that the minimum threshold, on the left sides of Eq. (\ref{eq:history}) \& (\ref{eq:ad-interaction}) i.e. taken as a minimum of `\({\eta _l^{\min }\left( {{\Psi _l}} \right)}\)' and `0', to respectively represent its \textit{weightage} as higher than the minimum weightage of present interests (and hence to show its importance in current \textit{targeting} components) or lower its dominating factor to slightly higher than `0'.

Subsequently, we have the following equivalent user profile, as a representation of \textit{weightage} of all the \textit{targeting criteria},:

\begin{equation}\label{eq:profile-weightages}
{I_g} = {\eta _l}\left( {{\Psi _l}} \right) + {\eta _m}\left( {{\Pi _m}} \right) + {\eta _n}\left( {{\Gamma _n}} \right)
\end{equation}

\subsection{Profile updating}
An important factor in \textit{ads targeting} is to track user's activity to find temporal changes in a user profile; hence, the \textit{profile} and \textit{ads targeting} is updated each time variations in user behavior is observed i.e. the \textit{targeting criteria} result in interests other than the existing set of interests. We note that following \emph{criteria} (but not limited to) is used to track changes in a use profile: \textit{installation/un-installation of an app} i.e. a user uses new set of \textit{apps} $S{'_a}$, which has no overlap with the existing set of \textit{apps} $S_a$, \textit{increase/decreases of use of an existing app}, \textit{start use of an `un-used' app}, \textit{interactions with ads}, and \textit{changes in web browsing history/searches}. Let \(\eta _l^{'\min }\) and \(\eta _l^{'\max }\) are new minimum and maximum threshold due to changes in \textit{apps usage}, then following conditions may hold (provided that the \textit{usage time} is \textit{i.i.d} i.e. \textit{independent and identical distributed}, with some unknown probability distribution):

\begin{eqnarray}
\eta _l^{'\max } > \eta _l^{\max };\,\eta _l^{'\min } > \eta _l^{\min } \to apps\,uninstalled \nonumber \\
\eta _l^{'\max } < \eta _l^{\max };\,\eta _l^{'\min } < \eta _l^{\min } \to new\,apps\,installed \nonumber \\
\eta _l^{'\max } < \eta _l^{\max };\,\eta _l^{'\min } < \eta _l^{\min } \to unused\,apps\,become\,used \\
\eta _l^{'\max } > \eta _l^{\max };\,\eta _l^{'\min } > \eta _l^{\min } \to used\,app\,become\,unused \nonumber
\end{eqnarray}

Based on the above constraints, we have the following equivalent constraint due to \textit{changes in-app activity}:

\begin{eqnarray}
\eta _l^{'\min }\left( {{\Psi _l}} \right) \ge \min \left\{ {\eta _l^{\min },\eta _l^{'\min }} \right\} \label{eq:equivalant-const-1}\\ 
\eta _l^{'\max }\left( {{\Psi _l}} \right) \ge \max \left\{ {\eta _l^{\max },\eta _l^{'\max }} \right\} \label{eq:equivalant-const-2}
\end{eqnarray}

In addition, such changes in \textit{browsing history/searches} and \textit{interactions with ads} can also be expressed with threshold limits, as explained in the next section.

\subsection{Profile evolution}\label{profile-evolution}
Furthermore, subsequent changes in a user profile (w.r.t. time $t$) are represented as \({C_t}\left( {{\eta _l}\left( {{\Psi _l}} \right)} \right)\); \({C_t}\left( {{\eta _m}\left( {{\Pi _m}} \right)} \right)\) and \({C_t}\left( {{\eta _n}\left( {{\Gamma _n}} \right)} \right)\) respectively for changes in \textit{apps usage}, \textit{web browsing history/searches} and \textit{interactions with ads}. Hence, Eq. (\ref{eq:profile-weightages}) can be re-written as:

\begin{equation}\label{eq:profile-change}
I_g^t = {C_t}\left( {{\eta _l}\left( {{\Psi _l}} \right)} \right) + {C_t}\left( {{\eta _m}\left( {{\Pi _m}} \right)} \right) + {C_t}\left( {{\eta _n}\left( {{\Gamma _n}} \right)} \right)
\end{equation}

Since the assigned \textit{weightages} should always be non-negative and bounded by maximum thresholds (i.e. maximum \textit{convergence point} in a specific period e.g. of every 24 hours), hence we need to make sure the following:

\begin{eqnarray}
0 < {C_t}\left( {{\eta _l}\left( {{\Psi _l}} \right)} \right) \le C_t^{\max }\left( {{\eta _l}\left( {{\Psi _l}} \right)} \right);\,\,\forall t,\forall l \in \left[ {1,\psi } \right] \label{eq:weightage-change1} \\
0 < {C_t}\left( {{\eta _m}\left( {{\Pi _m}} \right)} \right) \le C_t^{\max }\left( {{\eta _m}\left( {{\Pi _m}} \right)} \right);\,\,\forall t \label{eq:weightage-change2} \\
0 < {C_t}\left( {{\eta _n}\left( {{\Gamma _n}} \right)} \right) \le C_t^{\max }\left( {{\eta _n}\left( {{\Gamma _n}} \right)} \right);\,\,\forall t \label{eq:weightage-change3} 
\end{eqnarray}

As mentioned earlier, the other profiling components, such as \textit{history/searches} and \textit{interactions with ads}, are also \textit{mapped} to profiling interests \({\Psi _l}\), to produce a unified profile with different dominating interest categories, e.g. for \textit{ad-interactions}: \(f:{\Gamma _n} \to {\Psi _l} = {S_{{K_a}}}\left\{ {{\Psi _l}} \right\}_{l = 1}^\psi \). We presume that all changes in user profiles are distributed with an unknown probability distribution and that these profile \textit{weightages} are deterministically bounded by a finite constant e.g. \(C_t^{\max }\left( {{\eta _m}\left( {{\Pi _m}} \right)} \right)\) for web history, so that (incorporated in user profile) at time slot `$t$', Eq. (\ref{eq:profile-change}) can be re-written, as a unified profiling interests, as:

\begin{equation}\label{eq:profile-final}
I_g^t = \sum\limits_{\forall t} {{C_t}\left( {{\eta _l}\left( {{\Psi _l}} \right)} \right)}; \,\,\,\forall l \in \left\{ {\left[ {1,\psi } \right],m,n} \right\}
\end{equation}

Similarly, the newly updated (or change in existing) interest categories are reflected in profile at `$t+1$' as: \(I_g^{t + 1} = I_g^t + {C_{t+1}}\left( {{\eta _l}\left( {{\Psi _l}} \right)} \right)\). For $n$ amount of time:

\begin{equation}\label{eq:profile-change-final}
0 < I_g^{t + 1} \le I_g^{t + n}\,\,\,\,\forall t
\end{equation}

\(I_g^{t + n}\) is the maximum \textit{convergence point} of a user profile i.e. the point where users are \textit{targeted} with the most relevant ads. We envisage that this information is regularly updated, e.g. once per 24 hours, which we call \textit{evolution threshold} i.e. the time required to \textit{evolve} profile's interests, and is used to reflect the updated profile and fetch associated \textit{targeted ads}:

\begin{equation}\label{eq:final-profile}
\mathop {\lim }\limits_{t \to \infty } \sum\limits_{\tau = 1}^t {I_g^\tau }
\end{equation}

Recall that profile at this stage, as \({t \to \infty }\), due to maximum profile convergence, it contains the dominating interests \({g_{k,l}}\) that are considered \textit{private} by user, and hence exposes user's privacy. Our goal is to design a control algorithm that jointly optimises user's \textit{privacy} (both profiling and \textit{app} usage activity at different time of the day/night, as detailed in next section) and \textit{cost} of recommended \textit{apps} (discussed in Section \ref{optimal-profiling}).

\subsection{Apps usage profile}\label{apps-usage-profiling}
Let \({U^t}\left( {{K_a}} \right)\) represents the \textit{apps} usage at time slot $t$ (during day/night, note, for simplicity we do not consider the \textit{app} usage time for individual \textit{apps}). The \textit{apps} usage behavior for every user is time-varying, e.g. some users play game/gambling \textit{apps} during the lunch time or early in bed when she gets up, or an employee might scroll through the stocks in a broker's application during lunch break, to be exposed to her employer. We note that the user's \textit{apps} usage activity can be intercepted from the connected network \cite{tchen2014} or through \textit{app}'s interactions with the advertising systems \cite{ullah2014characterising, ullah2020privacy}. Let \(\mathop {\lim }\limits_{t \to \infty } \sum\limits_{\tau = 1}^t {{U^\tau }\left( {{K_a}} \right)} \) denotes \textit{app}'s usage at different time slots. The \textit{app}'s usage time varies during the 24 hours for each user, which also exposes user's privacy (irrespective of the category of \textit{app}'s usage) in terms of use of \textit{apps} during various times of the day/night.

\section{Optimal Privacy Preserving Profiling} \label{optimal-profiling}

Based on the above requirements for user profiling, we now study optimal privacy-preserving profile management that is \textit{cost effective}, \textit{preserve user's privacy due to profiling interests} and the \textit{apps usage behavior}, in addition to, user's \textit{web history/searches} and \textit{interactions with ads}. 

\subsection{Proposed framework}
Figure \ref{system-model} presents detailed overview of the proposed framework, introducing changes to `User Environment' that implements local user profiling both for \textit{Context profile} and \textit{Interests profiles} and further protects their privacy, presented in Section \ref{protect-sensitive-attr}, similarly, preserves user's privacy for user's \textit{apps usage behavior} during day/night time, detailed in Section \ref{prtect-apps-usage}, implemented via `System App'. We implement these various functionalities via a Proof of Concept (PoC) mobile \textit{app}; detailed discussion is given in Section \ref{sec:obfusSystem}. The `System App' also implements this framework that jointly optimises the profiling process and preserves user privacy in a cost-effective way, as detailed in Sections \ref{objective-function}, \ref{problem-formulation}, and \ref{problem-relaxation}. We suggest that this framework can be integrated into \texttt{AdMob SDK} since the current ad ecosystem carries out user profiling and \textit{targeted ads} via \texttt{SDK}s, which will require \texttt{SDK} modifications.

\begin{figure*}[h]
\begin{center}
\includegraphics[scale=0.58]{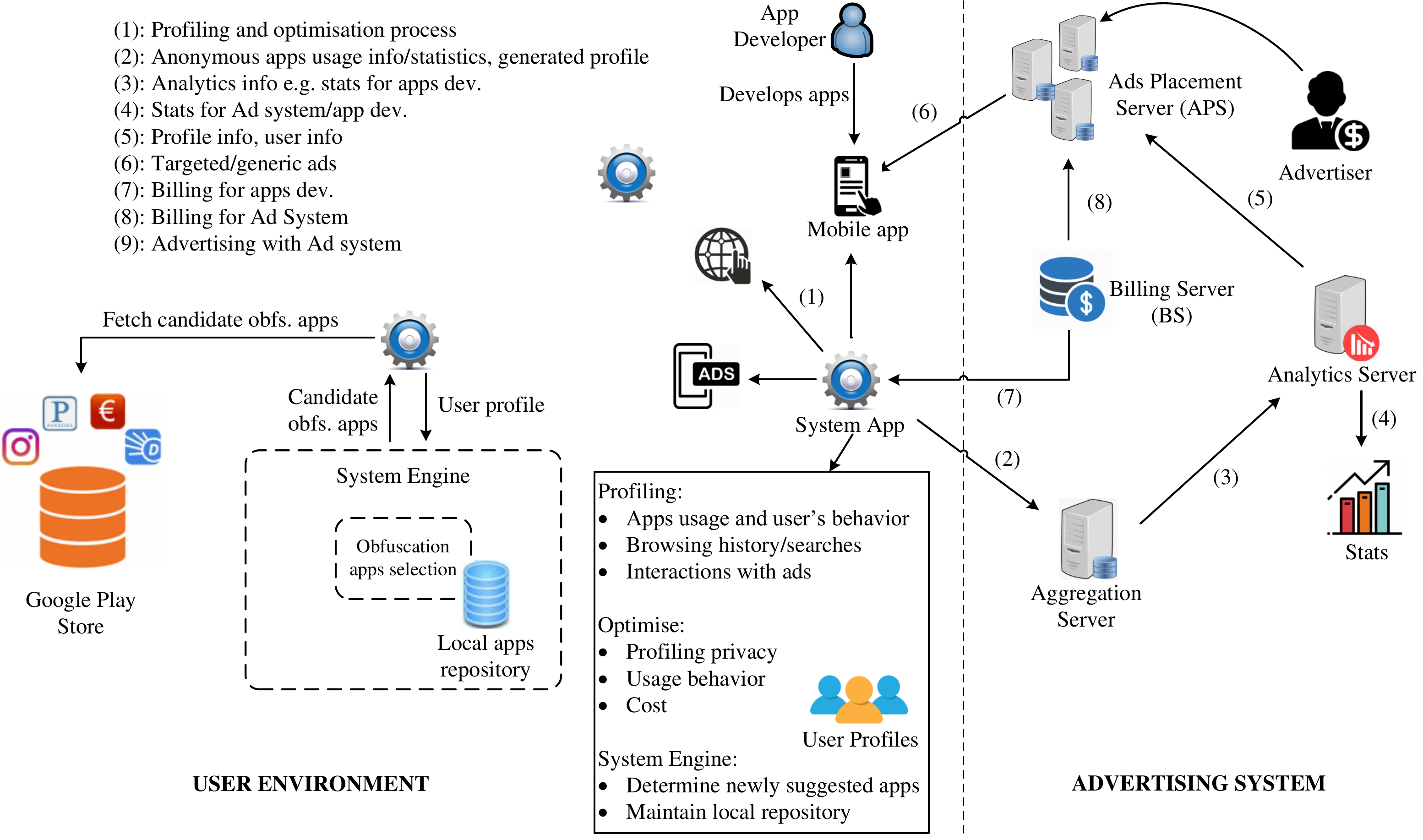}
\caption{The proposed advertising system with changes in user and advertising environments. \label{system-model}}
\end{center}
\end{figure*}

\subsection{Protecting sensitive profiling interests}\label{protect-sensitive-attr}
To protect the private interest categories i.e. sensitive to users, we select various other \textit{apps} based on \textit{similarity} metric \cite{ullah2014profileguard}\footnote{Detailed discussion over profile \textit{obfuscation} to achieve user privacy in an \textit{app}-based user profiling, using various \textit{obfuscation} strategies, can be found in \cite{ullah2014profileguard}.} to reduce the dominating private interest categories present in a user profile. This metric is calculated based on \textit{app} keywords \({\kappa _{i,j}}\) using the $tf-idf$ (\textit{cosine similarity}) metric \cite{ullah2014characterising}. This strategy is not metric-specific hence other similarity metrics e.g., \textit{jaccard index}, can also be used.

The newly selected candidate \textit{apps}, which we call the \textit{obfuscation apps} (note that in this work, we interchangeably use both `recommended apps' and `obfuscation apps'), are selected (and run for specific amount of time, as described in Eq. (\ref{first-scenario}), (\ref{obfuscation-weightage}), and (\ref{third-scenario})) from \textit{apps} categories $\Phi _j$, $j=1,..., \phi $ other than the private category $\Phi _p$; $\Phi _p$ is considered private by users that they want to protect. The recommended set $S_o$ comprises those \textit{apps} with highest similarity to the existing set of \textit{apps} i.e. $S_a$. In addition, the user may protect any number of private interests categories $\Psi_p$, $p = 1, \cdots ,\Omega $ and $\Omega $ is the set of interests categories that are \textit{private} to the user. Furthermore, we presume that the selected \textit{app}(s) will always generate the profiling interests other than the private profiling interest(s) i.e. \(l' \ne l\).

We assign weightage \({\eta _{l'}}\left( {{\Psi _{l'}}} \right)\) to the newly generated profiling interests, \(I_g^{'t}\), in order to reflect its effect over \textbf{1}. \textit{privacy} (i.e. disruption in user profile) and \textbf{2}. \textit{targeted ads} (disruption in receiving \textit{targeted ads} based on private profiling interests, exposed to ad/analytics networks):

\begin{equation}
\eta _l^{\min }\left( {{\Psi _l}} \right) > {\eta _{l'}}\left( {{\Psi _{l'}}} \right) > 0
\end{equation}

\textit{Cost} is usually defined as the ratio of obfuscating to original data \cite{parra2010privacy}. We further elaborate on \textit{Cost} in Section \ref{subsection:cost}. Recall that, beyond privacy protection, the use of these recommended \textit{apps} is used to protect user privacy of \textit{apps usage behavior} e.g., during the 24 hours period.

\subsection{Control objective for apps usage behavior}\label{prtect-apps-usage}
As mentioned in Section \ref{apps-usage-profiling}, intuitively speaking, in order to achieve user privacy for \textit{apps} usage at different time slots $t$ of the day, the \textit{apps} usage profile \({U^t}\left( {{K_a}} \right)\) needs to be `flatten' as \({t \to \infty }\), as much as possible, by running additional (recommended \textit{apps}, as described in above section) \textit{apps} i.e. \(U{^{'t}}\left( {{K_a}} \right)\). Subsequently, the profile becomes:

\begin{equation}\label{eq:obfus-profile}
I_g^{'t} = I_g^t + {U^{'t}}\left( {{K_a}} \right)
\end{equation}

Let \(\overline {{U^t}\left( {{K_a}} \right)} = \mathop {\lim }\limits_{t \to \infty } \frac{1}{t}\sum\limits_{\tau = 1}^t {{U^\tau }\left( {{K_a}} \right)} \) represents the average \textit{apps} usage for a user \({{K_a}}\), alternatively \(\overline {{U^t}\left( {{K_a}} \right)} = \overline {{C_t}\left( {{\eta _l}\left( {{\Psi _l}} \right)} \right)} ,\forall l \in \left[ {1,\psi } \right]\). In real time, at different time slots, the \textit{apps} usage needs to be controlled with as little deviation from \(\overline {{C_t}\left( {{\eta _l}\left( {{\Psi _l}} \right)} \right)} \) as possible. Note that this is also applicable to the interests in \({I_g^t}\) generated via \textit{browsing history/searches} and \textit{interactions with ads}, since, as mentioned earlier that \({\Pi _m} \to {\Psi _l}\) and \({\Gamma _n} \to {\Psi _l}\), in addition, using both these activities, the user devices interact with the ad/analytic networks for tracking user's activity along with their (device's) usage behavior.

Subsequently, the control objective is to minimise the variance of \(\overline {{U^t }\left( {{K_a}} \right)} \), i.e.:

\begin{equation}\label{eq:apps-usage-behv}
\mathop {\lim }\limits_{t \to \infty } \frac{1}{t}\sum\limits_{\tau = 1}^t {\mathbb{E} \left\{ {{{\left( {I_g^\tau - \overline {U{'^\tau }\left( {{K_a}} \right)} } \right)}^2}} \right\}}
\end{equation}

The privacy protection in \textit{apps usage behavior}, in addition to, \textit{preserve user's privacy due to profiling interests}, can be achieved by using a few other suggested \textit{apps}, described in the next section, for our proposed scenario. 

\subsection{Objective function}\label{objective-function}
In this paper, we jointly optimise the user's privacy due to \textit{profiling interests}, the \textit{cost of running obfuscation apps}, and the \textit{apps usage behavior}. The objective function can be expressed as:

\begin{equation}
\mathop {\lim }\limits_{t \to \infty } \sum\limits_{\tau = 1}^t {\left( \begin{array}{l}
\underbrace {{C^\tau }{\eta _{l'}}\left( {{\Psi _{l'}}} \right)}_{Cost\,of\,running\,obfs.\,apps} + \underbrace {\beta \left( {I_g^\tau + \eta _{l'}^\tau \left( {{\Psi _{l'}}} \right)} \right)}_{Profiling\,privacy} + \\
\underbrace {{\beta \left( {I_g^\tau - \overline {{U{'^\tau }}\left( {{K_a}} \right)} } \right)}^2}_{Apps\,usage\,privacy}
\end{array} \right)}
\end{equation}

Here, the $\beta$ parameter is selected by user in order to achieve a trade-off between user \textit{privacy} and \textit{targeted ads}. Note that the selection of this parameter affects the \textit{targeted ads} as a result of disruption in a user profile. We describe the following various scenarios for introducing (along with the number of \textit{apps}) recommended \textit{apps}:

\begin{eqnarray}
0 < {\eta _{l'}}\left( {{\Psi _{l'}}} \right) \le \eta _l^{\min }\left( {{\Psi _l}} \right) \label{first-scenario} \\
\eta _l^{\min }\left( {{\Psi _l}} \right) \le {\eta _{l'}}\left( {{\Psi _{l'}}} \right) \le \eta _l^{\max }\left( {{\Psi _l}} \right) \label{obfuscation-weightage} \\
{\eta _{l'}}\left( {{\Psi _{l'}}} \right) \ge \eta _l^{\max }\left( {{\Psi _l}} \right) \label{third-scenario}
\end{eqnarray}

\begin{figure}[h]
\begin{center}
\includegraphics[scale=0.46]{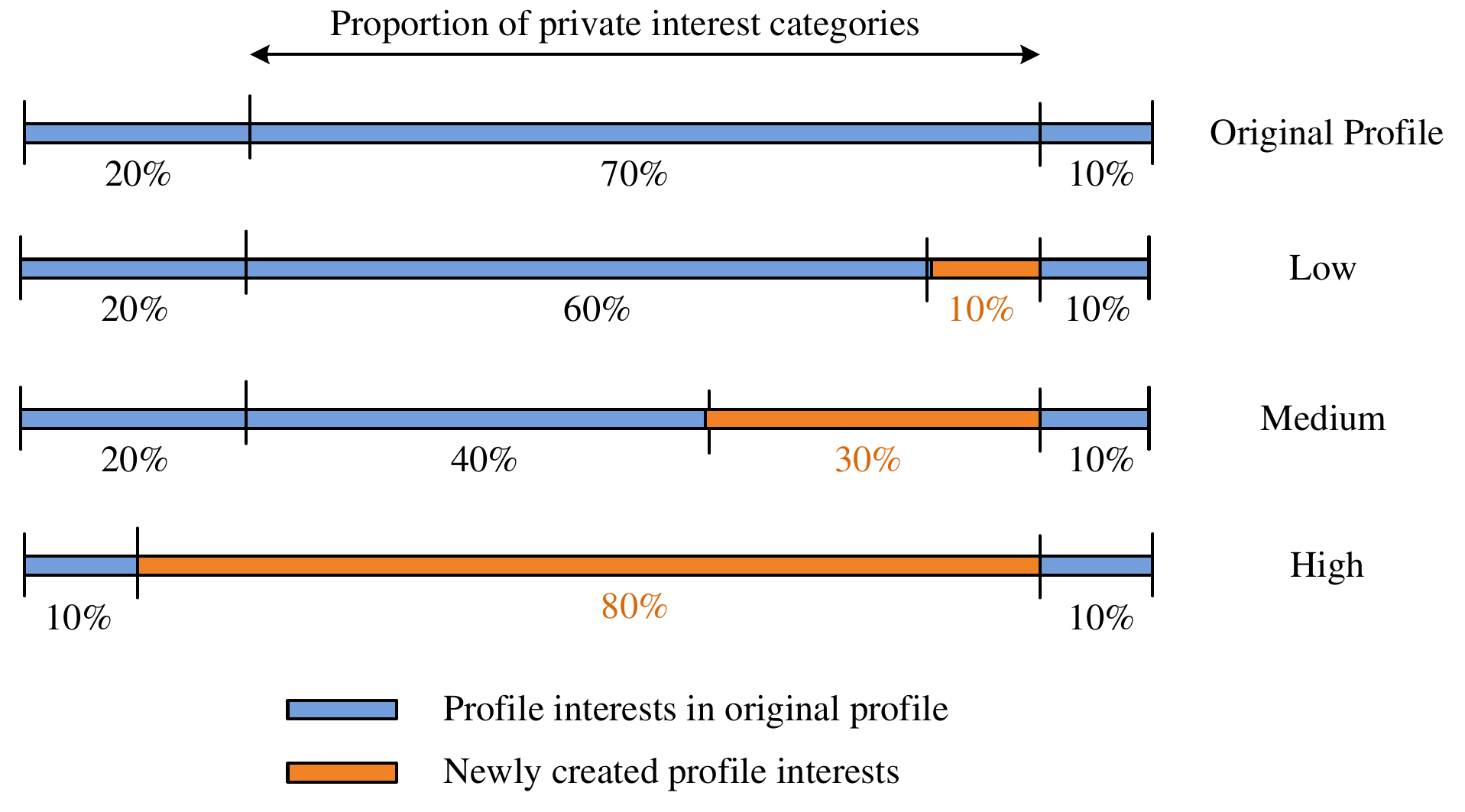}
\caption{Trade-off between \textit{privacy} and \textit{targeted ads} achieved via various obfuscation scenarios of \textit{Low}, \textit{Medium}, and \textit{High} profile (interests), respectively introducing 10\%, 30\%, and 80\% disruption in a user profile. \label{emperical-example}}
\end{center}
\end{figure}

The first scenario (i.e. Eq. (\ref{first-scenario})) introduces new \textit{obfuscation apps} that would introduce minimum disruption in a user profile, hence, achieves lower \textit{privacy} and attracts higher \textit{targeted ads}, as opposed to the last scenario, which introduces highest disruption in a user profile i.e., achieves higher \textit{privacy} and attracts less relevant \textit{targeted ads}. On the other hand, the middle scenario introduces medium disruption in a user profile and achieves a balance between user \textit{privacy} and \textit{targeted ads}. An empirical example for various scenarios is given in Figure \ref{emperical-example}. We envisage that this scenario (\ref{obfuscation-weightage}) will further introduce medium operating \textit{cost} of the selected \textit{obfuscation apps} i.e. \({C^\tau }{\eta _{l'}}\left( {{\Psi _{l'}}} \right)\).

\subsection{Problem formulation}\label{problem-formulation}
The optimal privacy preserving user profiling and \textit{targeted ads} can be formulated as a dynamic optimisation problem:

\begin{equation}\label{eq:prob-formulation}
min \mathop {\lim }\limits_{t \to \infty } \sum\limits_{\tau = 1}^t {\left( \begin{array}{l}
{C^\tau }{\eta _{l'}}\left( {{\Psi _{l'}}} \right) + \beta \left( {I_g^\tau + \eta _{l'}^\tau \left( {{\Psi _{l'}}} \right)} \right) + \\
\beta {\left( {I_g^\tau - \overline {{U{'^\tau } }\left( {{K_a}} \right)} } \right)}^2
\end{array} \right)}
\end{equation}

s.t. constants: (\ref{eq:weightage-change1}), (\ref{eq:weightage-change2}), (\ref{eq:weightage-change3}), (\ref{eq:profile-change-final}), (\ref{obfuscation-weightage}).

An important challenge to solve this optimisation problem is to know the user's temporal behavior as a combined activity of `Web \& App' i.e. the lack of knowledge of time-varying updates in a user profile. This change in temporal behavior is affected by the profiling derived by \textit{app's usage} Eq. (\ref{eq:weightage-change1}), \textit{browsing history/searches} Eq. (\ref{eq:weightage-change2}), and \textit{interactions with ads} Eq. (\ref{eq:weightage-change3}). This problem would become even more challenging when there is an irregular activity observed by a user e.g. high app's/web usage during weekends etc. Hence, we develop an optimal control algorithm, using \textit{Lyapunov optimization}, for identifying updates in user profiles as a result of communication requests between the mobile device and the advertising system (analytics) entities, see Section \ref{online-control-algo}. Recall that ad/analytics \texttt{SDK}s enable these requests for tracking and profiling users for their `Web \& App' activities.

\subsection{Problem relaxation}\label{problem-relaxation}
To solve the optimisation problem (\ref{eq:prob-formulation}), we consider its relax version using (\ref{eq:profile-final}) to relax constraints (\ref{eq:weightage-change1}), (\ref{eq:weightage-change2}), and (\ref{eq:weightage-change3}). The average expected change in a user profile \({I_g^t}\) is given by:

\begin{equation}\label{eq:relaxed-const}
\overline {I_g^t} = \mathop {\lim }\limits_{t \to \infty } \frac{1}{t}\sum\limits_{\tau = 1}^t {\mathbb{E} \left\{ {{C_t}\left( {{\eta _l}\left( {{\Psi _l}} \right)} \right)} \right\}}
\end{equation}

From Eq. (\ref{eq:profile-final}), it is clear that the profile \textit{evolves} over time e.g. over time slot $t$ and $t+1$ we have \(I_g^t = {C_{t + 1}}\left( {{\eta _l}\left( {{\Psi _l}} \right)} \right) + {C_t}\left( {{\eta _l}\left( {{\Psi _l}} \right)} \right)\). Hence, we take its expected values on both sides of Eq. (\ref{eq:profile-final}) and equate it to Eq. (\ref{eq:relaxed-const}), we have:

\begin{equation}\label{eq:profile-infinite}
\sum\limits_{\tau = 1}^t {\mathbb{E}\left\{ {{C_t}\left( {{\eta _l}\left( {{\Psi _l}} \right)} \right)} \right\}} = {C_{t - 1}}\left( {{\eta _l}\left( {{\Psi _l}} \right)} \right) + {C_t}\left( {{\eta _l}\left( {{\Psi _l}} \right)} \right)
\end{equation}

Recall that \({C_t}\left( {{\eta _l}\left( {{\Psi _l}} \right)} \right)\) is the initial \textit{weightages} of profiling interests as the profile \textit{evolves}. Similarly, as mentioned earlier that, these changes are bounded by finite $min$ and $max$ bounds i.e.:

\begin{equation}\label{eq:cost-bounds}
C_t^{\min }\left( {{\eta _l}\left( {{\Psi _l}} \right)} \right) \le {C_t}\left( {{\eta _l}\left( {{\Psi _l}} \right)} \right) \le C_t^{\max }\left( {{\eta _l}\left( {{\Psi _l}} \right)} \right)
\end{equation}

Divide both sides of Eq. (\ref{eq:profile-infinite}) by $t$ and take \({t \to \infty }\). Following:

\begin{equation}\label{eq:zero-change}
\overline {I_g^t} = 0
\end{equation}

Consequently, we have the following relaxed version of objective function:

\begin{equation}\label{eq:relaxed1}
min \mathop {\lim }\limits_{t \to \infty } \sum\limits_{\tau = 1}^t {\left( \begin{array}{l}
{C^\tau }{\eta _{l'}}\left( {{\Psi _{l'}}} \right) + \beta \left( {I_g^\tau + \eta _{l'}^\tau \left( {{\Psi _{l'}}} \right)} \right) + \\
\beta {\left( {I_g^\tau - \overline {{U{'^\tau } }\left( {{K_a}} \right)} } \right)}^2
\end{array} \right)}
\end{equation}

s.t. constants: (\ref{eq:profile-final}), (\ref{eq:profile-change-final}), (\ref{obfuscation-weightage}).

Now the main challenge of solving objective (\ref{eq:relaxed1}) is to minimise the variance of \(\overline {{U^t}\left( {{K_a}} \right)} \) in order to protect user's \textit{app's usage behavior} i.e. to know all the future usage of \textit{apps} and hence suggest the (automatic) use of recommended \textit{obfuscation apps} i.e. \(\overline {{U{'^t }}\left( {{K_a}} \right)} \). Similar to Eq. (\ref{eq:zero-change}), we can also show that \({\overline {{U^t}\left( {{K_a}} \right)} = 0}\) i.e. to prove that decision variable is independent of \({{U^t}\left( {{K_a}} \right)}\). Hence, the above optimisation problem can be solved without the information of \(\overline {{U{'^t }}\left( {{K_a}} \right)} \) at different time slots $t$, as detailed below.

\bigskip

\paragraph*{Proof} From Eq. (\ref{eq:apps-usage-behv}), it can be shown that:

\begin{equation}\label{eq:const-apps-usage}
\mathop {\lim }\limits_{t \to \infty } \frac{1}{t}\sum\limits_{\tau = 1}^t {\mathbb{E} \left\{ {\left( {{{\left( {\overline {U{'^\tau }\left( {{K_a}} \right)} } \right)}^2} - 2I_g^\tau \overline {U{'^\tau }\left( {{K_a}} \right)} } \right)} \right\}}
\end{equation}

The above equation is independent of the choice of \({\overline {U{'^\tau }\left( {{K_a}} \right)} }\), hence:

\begin{equation}\label{eq:const-apps-usage2}
\mathop {\lim }\limits_{t \to \infty } \frac{1}{t}\sum\limits_{\tau = 1}^t {\mathbb{E} \left\{ {I_g^\tau } \right\}}
\end{equation}

As shown in Eq. (\ref{eq:obfus-profile}), sum over all $t$, take the expectation of both sides and take \({t \to \infty }\), we have the following:

\[\mathop {\lim }\limits_{t \to \infty } \frac{1}{t}\sum\limits_{\tau = 1}^t {\mathbb{E} \left\{ {I_g^{'\tau }} \right\}} = \mathop {\lim }\limits_{t \to \infty } \frac{1}{t}\sum\limits_{\tau = 1}^t {\mathbb{E} \left\{ {I_g^\tau } \right\}} + \]

\begin{equation}\label{eq:const-apps-usage3}
\mathop {\lim }\limits_{t \to \infty } \frac{1}{t}\sum\limits_{\tau = 1}^t {\mathbb{E} \left\{ {\overline {U{'^\tau }\left( {{K_a}} \right)} } \right\}}
\end{equation}

\[ = \overline {I_g^t} + \overline {{U^{'t}}\left( {{K_a}} \right)} \]

From Eq. (\ref{eq:zero-change}), we conclude:

\begin{equation}\label{eq:const-apps-usage3}
\mathop {\lim }\limits_{t \to \infty } \frac{1}{t}\sum\limits_{\tau = 1}^t {\left\{ {I_g^{'\tau }} \right\}} = \overline {{U^{'t}}\left( {{K_a}} \right)}
\end{equation}

\begin{flushright}
$\blacksquare$
\end{flushright}

The relax version of our objective function in Eq. (\ref{eq:relaxed1}) can be re-written as:

\begin{equation}\label{eq:relaxed2}
\mathop {\lim }\limits_{t \to \infty } \sum\limits_{\tau = 1}^t {\left( {{C^\tau }{\eta _{l'}}\left( {{\Psi _{l'}}} \right) + \beta \left( {I_g^\tau + \eta _{l'}^\tau \left( {{\Psi _{l'}}} \right)} \right) + \beta \left( {I_g^\tau } \right)} \right)}
\end{equation}

Note that this is an optimised version of (\ref{obfuscation-weightage}) i.e. the scenario of introducing medium disruption in a user profile and it achieves balance between user \textit{privacy} and \textit{targeted ads}. The other scenarios of low (\ref{first-scenario}) and high (\ref{third-scenario}) profile disruption will have no effect over the (\ref{eq:relaxed2}) except low and high costs and trade-off between \textit{privacy} and \textit{targeted ads}.

\section{Optimal Control Algorithm} \label{online-control-algo}

We design a control algorithm for identifying communication requests between the mobile device and the analytics server within an advertising ecosystem, to achieve an optimal solution to (\ref{eq:prob-formulation}) for identifying time varying updates in user profiles. Let \(R_l^t\) represents an advertising request reported at $t$ via ad/analytics \texttt{SDK} i.e. either ad request for display/ad click in \textit{apps}/web or web searches/history; recall that a request may or may not introduce $l$ ($l$ is profiling interest category introduced during profile updating or \textit{evolution}) in a user profile \(I_g^t\), alternatively, a change in user profile with an addition of \({C_t}\left( {{\eta _l}\left( {{\Psi _l}} \right)} \right)\) during the profile \textit{evolution}, as described in Section \ref{profile-evolution}. Hence:

\begin{equation}\label{ad-request-tracking}
R_l^t = I_g^t - C_t^{\min }\left( {{\eta _l}\left( {{\Psi _l}} \right)} \right)
\end{equation}

Subsequently, the \(R_l^t\) is a shifted version of \(I_g^t\) and can be described as:

\begin{equation}\label{ad-request-shifted}
R_l^{t + 1} = I_g^t + R_l^t
\end{equation}

\subsection{Lyapunov optimisation}
For $n$ requests, we have \(R_l^t = R_l^1, \cdots ,R_l^n\), following, the quadratic \textit{Lyapunov function} for each $t$ is given as:

\begin{equation}\label{lyapunov-function}
Q\left( {R_l^t} \right) = \sum\limits_{\tau = 1}^t {{{\left( {R_l^\tau } \right)}^2}}
\end{equation}

Similarly, the corresponding \textit{Lyapunov drift} i.e. \textit{drift-plus-penalty}, can be defined as:

\begin{equation}\label{lyapunov-drift}
\Delta R_l^t = \mathbb{E} \left\{ {Q\left( {R_l^{t + 1}} \right) - Q\left( {R_l^t} \right)|I_g^t} \right\}
\end{equation}

To stabilize the upper bound on Eq. (\ref{eq:relaxed2}), while effective profiling process i.e. the number of interests drawn, to minimise cost of recommended \textit{apps}, and to minimise the variance of \(\overline {{U^t }\left( {{K_a}} \right)} \) for preserving user privacy of \textit{apps usage} behavior, the control algorithm can be designed to minimise the following \textit{drift-plus-penalty} on each time slot $t$:

\begin{equation}\label{lyapunov-panalty}
\Delta R_l^t + Vp\left( t \right)
\end{equation}

\sloppy Here, \(p\left( t \right)\) is the penalty for minimising the upper bound on Eq. (\ref{eq:relaxed2}) i.e. minimise \(p\left( t \right) = \mathbb{E} \left\{ {{C^t}{\eta _{l'}}\left( {{\Psi _{l'}}} \right) + \beta \left( {I_g^t + \eta _{l'}^t\left( {{\Psi _{l'}}} \right)} \right) + \beta I_g^t} \right\}\), \(V > 0\) is a non-negative parameter chosen as a desired effect of performance of trade-off over the objective function. This approach does not require the knowledge of all future events i.e. the lack of knowledge of time varying updates in a user profile and the \textit{apps} usage behavior. 

\paragraph*{Lemma} The upper \textit{drift} bound. For the control policy that satisfies the constraints on Eq. (\ref{eq:relaxed2}), we have the following \textit{drift-plus-penalty} condition holds:

\begin{equation}\label{lyapunov-panalty-condition}
\mathbb{E} \left\{ {\Delta R_l^t + Vp\left( t \right)|Q\left( {R_l^t} \right)} \right\} \le B + Vp'\left( t \right) - \varepsilon \sum\limits_{\tau = 1}^t {\left( {R_l^\tau } \right)}
\end{equation}

$\varepsilon$ is similar to \(\beta \) and can be controlled to achieve the average \textit{apps} usage time to preserve privacy of user's \textit{app usage} behavior and to achieve a trade-off between user \textit{privacy} and \textit{targeted ads}. \(p'\left( t \right)\) is the desired \textit{target} for the time average of \(p\left( t \right)\). $B$ can be defined as:

\begin{equation}\label{lyapunov-panalty-constant-B}
B = \frac{1}{2}\max \left\{ {C_t^{\max }\left( {{\eta _l}\left( {{\Psi _l}} \right)} \right),C_t^{\min }\left( {{\eta _l}\left( {{\Psi _l}} \right)} \right)} \right\}
\end{equation}

\begin{equation}\label{lyapunov-panalty-condition-further}
\frac{1}{t}\sum\limits_{\tau = 1}^t \mathbb{E} {\left\{ {p(\tau )} \right\}} \le p'\left( t \right) + \frac{B}{V} + \frac{{\mathbb{E} \left\{ {R_l^1} \right\}}}{{Vt}}
\end{equation}

\bigskip
\paragraph*{Proof} Taking expectations of both sides of above \textit{drift-plus-penalty}, we have:

\begin{equation}\label{lyapunov-panalty-proof1}
\mathbb{E} \left\{ {\Delta R_l^t} \right\} + V \mathbb{E} \left\{ {p\left( t \right)} \right\} \le B + Vp'\left( t \right) - \varepsilon \sum\limits_{\tau = 1}^t {\left( {R_l^\tau } \right)}
\end{equation}

\begin{eqnarray}
\mathbb{E} \left\{ {R_l^t} \right\} - \mathbb{E} \left\{ {R_l^1} \right\} + V\sum\limits_{\tau = 1}^t {\mathbb{E} \left\{ {p(\tau )} \right\}} \le \nonumber \\
 \left( {B + Vp'\left( t \right)} \right)t - \varepsilon \sum\limits_{\tau = 0}^t {\mathbb{E} \left\{ {Q\left( {R_l^\tau } \right)} \right\}} - \mathbb{E} \left\{ {R_l^1} \right\} \nonumber \\
 + V\sum\limits_{\tau = 0}^t {\mathbb{E} \left\{ {p\left( \tau \right)} \right\}} \le \left( {B + Vp'\left( t \right)} \right)t \nonumber
\end{eqnarray}

\begin{equation}\label{lyapunov-panalty-proof2}
V\sum\limits_{\tau = 1}^t {\mathbb{E} \left\{ {p(\tau )} \right\}} \le p'\left( t \right)Vt + Bt + \mathbb{E} \left\{ {R_l^1} \right\}
\end{equation}

Dividing by $Vt$ and rearranging terms proves the bound on average penalty, concludes the proof. \,\,\,\,\,\,\,\,\,\,\,\,\,\,\,\,\,\,\,\,\,\,\,\,\,\,\,\,\,\,\,\,\,\,\,\,\,\,\,\,\,\,\,\, $\blacksquare$

\begin{algorithm*}
\caption{\hl{The control algorithm for joint optimisation of privacy and cost of in-app mobile user profiling and targeted ads.}}
\begin{algorithmic}[1]
\State \textbf{Initialization}
\State The initial user profile with various profiling interests form Eq. (\ref{eq:profile-final}) i.e., $I_g^t = \sum\limits_{\forall t} {{C_t}\left( {{\eta _l}\left( {{\Psi _l}} \right)} \right)}; \,\,\,\forall l \in \left\{ {\left[ {1,\psi } \right],m,n} \right\}$
\State The \textit{Interest profile} at $t+n$ i.e., at the convergence point from Eq. (\ref{eq:profile-change-final}) i.e., 
$0 < I_g^{t + 1} \le I_g^{t + n}\,\,\,\,\forall t$
\State The initial \textit{Context profile}, Eq. (\ref{initial-app-profile}) ${K_a} = \left\{ {\left\{ {{a_{i,j}},{\Phi _j}} \right\}:{a_{i,j}} \in {S_a}} \right\}$
\State The converged \textit{Context profile} for \(\mathop {\lim }\limits_{t \to \infty } \sum\limits_{\tau = 1}^t {{U^\tau }\left( {{K_a}} \right)} \) i.e., $0 < K_a^{t + 1} \le K_a^{t + n}\,\,\,\,\forall t$
\State The Eq. (\ref{eq:obfus-profile}) $I_g^{'t} = I_g^t + {U^{'t}}\left( {{K_a}} \right)$ where \(\overline {{U^t}\left( {{K_a}} \right)} = \mathop {\lim }\limits_{t \to \infty } \frac{1}{t}\sum\limits_{\tau = 1}^t {{U^\tau }\left( {{K_a}} \right)} \)
\State Take the shifted version of \(I_g^t\) i.e., \(R_l^t\) from Eq. (\ref{ad-request-shifted}) $R_l^{t + 1} = I_g^t + R_l^t$

\textbf{Evaluate the following over the time}
\State \textit{Lyapunov drift} using Eq. (\ref{lyapunov-drift})
$\Delta R_l^t = \mathbb{E} \left\{ {Q\left( {R_l^{t + 1}} \right) - Q\left( {R_l^t} \right)|I_g^t} \right\}$ and Eq. (\ref{lyapunov-panalty}) $\Delta R_l^t + Vp\left( t \right)$
\State The average \textit{penalty} using Eq. (\ref{lyapunov-panalty-proof2}) $V\sum\limits_{\tau = 1}^t {\mathbb{E} \left\{ {p(\tau )} \right\}} \le p'\left( t \right)Vt + Bt + \mathbb{E} \left\{ {R_l^1} \right\}$
\State The current state of $I_g^t$ and evaluate the problem using Eq. (\ref{lyapunov-panalty-drift-minimise})
$min \, R_l^t + V\left( {{C^t}{\eta _{l'}}\left( {{\Psi _{l'}}} \right) + \beta \left( {I_g^t + \eta _{l'}^t\left( {{\Psi _{l'}}} \right)} \right) + \beta I_g^t} \right)$ in the presence of constraint defined in Eq. (\ref{lyapunov-panalty-drift-minimise}).
\State For the \textit{stable state} of a profile, update the profile using Eq. (\ref{lyapunov-panalty-drift-minimise-case1}), $V\left( {{C^t}{\eta _{l'}}\left( {{\Psi _{l'}}} \right) + \beta \eta _{l'}^t\left( {{\Psi _{l'}}} \right)} \right)$.
\State For profile \textit{development and evolution states}, update the profile using Eq. (\ref{lyapunov-panalty-drift-minimise-case2}), $p\left( {R_l^t} \right) + V\left( {{C^t}{\eta _{l'}}\left( {{\Psi _{l'}}} \right) + \beta \left( {I_g^t + \eta _{l'}^t\left( {{\Psi _{l'}}} \right)} \right) + \beta I_g^t} \right)$. 
\State For various scenarios of minimum (\({p^{\min }}\left( {R_l^t} \right)\)), average (\({p^{avg}}\left( {R_l^t} \right)\)), and maximum (\({p^{\max }}\left( {R_l^t} \right)\)) values of \(p\left( {R_l^t} \right)\), update the value of $R_l^t$ using Eq. (\ref{ad-request-shifted}).
\end{algorithmic} \label{algorithm}
\end{algorithm*}

\subsection{Control algorithm}\label{control-algorithm}
The main objective of the control algorithm is to minimise the \textit{drift-plus-penalty} bound subject to the constraints of (\ref{eq:relaxed2}) at each time slot $t$. Detailed description of the control algorithm can be found in Algorithm \ref{algorithm}. This control algorithm optimally selects the minimum and maximum bounds over the selection of obfuscation \textit{apps}, the rate at which these \textit{apps} needs to run according to the variation in $R_l^t $ and by observing the current states of \({I_g^t}\) i.e., a solution to the following optimisation problem:

\begin{equation}\label{lyapunov-panalty-drift-minimise}
min \, R_l^t + V\left( {{C^t}{\eta _{l'}}\left( {{\Psi _{l'}}} \right) + \beta \left( {I_g^t + \eta _{l'}^t\left( {{\Psi _{l'}}} \right)} \right) + \beta I_g^t} \right)
\end{equation}

s.t. \begin{eqnarray}
C_t^{\min }\left( {{\eta _l}\left( {{\Psi _l}} \right)} \right) \le {C_t}\left( {{\eta _l}\left( {{\Psi _l}} \right)} \right) \le C_t^{\max }\left( {{\eta _l}\left( {{\Psi _l}} \right)} \right) \nonumber \\
R_l^t = I_g^t - C_t^{\min }\left( {{\eta _l}\left( {{\Psi _l}} \right)} \right) \nonumber \\
I_g^t = \sum\limits_{\forall t} {{C_t}\left( {{\eta _l}\left( {{\Psi _l}} \right)} \right)} {\mkern 1mu} {\mkern 1mu} {\mkern 1mu} \forall l \in \left\{ {\left[ {1,\psi } \right],m,n} \right\} \nonumber
\end{eqnarray}

The performance of this algorithm is to achieve minimised objective when changes occur in the profiling process at $t$ i.e. the \textit{stable} and during the \textit{profile development} and \textit{profile evolution} processes, in order to solve (\ref{lyapunov-panalty-drift-minimise}) as a mixed-integer non-linear programming optimisation problem. The following cases:

\bigskip
\paragraph*{Stable State} Recall that during this state, no changes occur in the profiling process i.e. \({{C_t}\left( {{\eta _l}\left( {{\Psi _l}} \right)} \right) = 0}\). Hence, the optimal value of (\ref{lyapunov-panalty-drift-minimise}) is evaluated to:

\begin{equation}\label{lyapunov-panalty-drift-minimise-case1}
V\left( {{C^t}{\eta _{l'}}\left( {{\Psi _{l'}}} \right) + \beta \eta _{l'}^t\left( {{\Psi _{l'}}} \right)} \right)
\end{equation}

Let \(p\left( {R_l^t} \right)\) tracks the advertising requests reported at $t$ via ad/analytics \texttt{SDK} during slot $t$, which is considered minimum i.e. \({p^{\min }}\left( {R_l^t} \right)\), during the \textit{stable} state, hence, the \({\eta _{l'}}\left( {{\Psi _{l'}}} \right)\) is selected as \({\eta _{l'}}\left( {{\Psi _{l'}}} \right) = \min \left\{ {0,\eta _l^{\min }\left( {{\Psi _l}} \right)} \right\}\).

\bigskip
\paragraph*{Profile Development/Evolution State} The profiling process speeds up during this state as a result of high interaction with the analytics servers. Let \({p^{\min }}\left( {R_l^t} \right)\), \({p^{avg}}\left( {R_l^t} \right)\), and \({p^{\max }}\left( {R_l^t} \right)\) respectively represents the minimum, average, and maximum; following we present various scenarios for calculating optimal value of (\ref{lyapunov-panalty-drift-minimise}):

\begin{itemize}
 \item For \({p^{\min }}\left( {R_l^t} \right)\): \begin{itemize}
      \item \({\eta _{l'}}\left( {{\Psi _{l'}}} \right) = \max \left\{ {0,\eta _l^{\min }\left( {{\Psi _l}} \right)} \right\}\)
     \end{itemize}
 \item For \({p^{avg}}\left( {R_l^t} \right)\): \begin{itemize}
     \item \({\eta _{l'}}\left( {{\Psi _{l'}}} \right) = \min \left\{ {\eta _l^{\min }\left( {{\Psi _l}} \right),\eta _l^{\max }\left( {{\Psi _l}} \right)} \right\}\)
    \end{itemize}
 \item For \({p^{\max }}\left( {R_l^t} \right)\): \begin{itemize}
     \item \sloppy \({\eta _{l'}}\left( {{\Psi _{l'}}} \right) = \max \left\{ {\eta _l^{\max }\left( {{\Psi _l}} \right),\eta _l^{\max }\left( {{\Psi _l}} \right) + \eta _{l'}^{\max }\left( {{\Psi _{l'}}} \right)} \right\}\)
    \end{itemize}
\end{itemize}

Given these values, the optimal objective can be calculated as:

\begin{equation}\label{lyapunov-panalty-drift-minimise-case2}
p\left( {R_l^t} \right) + V\left( {{C^t}{\eta _{l'}}\left( {{\Psi _{l'}}} \right) + \beta \left( {I_g^t + \eta _{l'}^t\left( {{\Psi _{l'}}} \right)} \right) + \beta I_g^t} \right)
\end{equation}

The lower values on (\ref{lyapunov-panalty-drift-minimise-case1}) and (\ref{lyapunov-panalty-drift-minimise-case2}) and the corresponding control values are calculated to evaluate the optimal values of our objective.

\section{Performance Measures} \label{perfor-analysis}

We now analyse the feasibility and performance of the proposed model using various evaluation metrics, in addition, we discuss the POC implementation of our proposed model. The applicability of the proposed system is discussed in Section \ref{applicability}.

\subsection{Evaluation metric}\label{sec:privacyMetric}
We define \textit{utility} and further elaborate the \textit{cost} of recommended \textit{apps} to provide insights on \textit{usability} of the recommended \textit{apps} usage. The authors in \cite{parra2010privacy} describe \textit{utility} as the success rate for removal of private query tags i.e. the magnitude suppression of original user preferences.

\subsubsection{Utility}
Let \({D\left( {{\eta _p}\left( {{\Psi _p}} \right)} \right)}\) is the dominating private interest category. We define \textit{utility} based on two components: first, from the effectiveness of privacy protection, we use as metric the level of reduction $R_p$ of \({{\eta _p}\left( {{\Psi _p}} \right)}\) of a selected private category ${\Psi _p}$ in an \textit{Interest profile}, achieved via recommended obfuscating apps, with \(\frac{{D\left( {{\eta _p}\left( {{\Psi _p}} \right)} \right)}}{{D\left( {{\eta _l}\left( {{\Psi _l}} \right)} \right)}}\). Here \({D\left( {{\eta _l}\left( {{\Psi _l}} \right)} \right)}\) is the new dominating ratio resulting from using \textit{apps} in $S_o$ ($S_o$ is the new set of \textit{apps} other than $S_a$; the selection of these \textit{apps} has been detailed in Section \ref{protect-sensitive-attr}).

Secondly, we introduce \textit{usability} $U_s$ of the selected obfuscating \textit{apps} that relates to the probability that a user would actually utilise these \textit{apps}, rather than just install and run it for privacy protection. The $U_s$ of an \textit{app} $a_o \in S_o$, in regards to a user with a specific \textit{Context profile}, as the ratio of similarity between this \textit{app} and any of the \textit{apps} in $S_a$ and the maximum similarity of any other \textit{app} from $\mathcal{A}$, not present in $S_a$ and \textit{apps} from the same set:

\begin{equation}\label{formula:usability}
\begin{array}{l}
{U_s} = min\left( {\frac{{sim\left( {{a_o},{a_{i,p}}} \right)}}{{sim\left( {{a_{q,r}},{a_{i,p}}} \right)}}} \right):\\
{a_{i,p}} \in {S_a},{\rm{ }}{a_{q,r}} \notin {\rm{ }}{S_a}
\end{array}
\end{equation}

Combining the two, the total \textit{utility} can then be calculated as: $U_T=R_p + U_s$.

\subsubsection{Cost and resource overhead}\label{subsection:cost}

The \textit{cost} and \textit{resource overhead}, can be considered as equivalent terms in the context of introducing new activities, which result in usage time and other resources e.g. battery usage, however, for the sake of clarity, we use two separate terms. As mentioned earlier, \textit{cost} $C$ is termed as a metric that relates to the reduction of overall usage time available (by use of obfuscating \textit{apps}) to original (non-obfuscating) \textit{apps}.

In the basic scenario where, the average introduction of recommended \textit{apps}, the usage of all \textit{apps} is uniformly distributed within a time period in correspondence to \({{U^t}\left( {{K_a}} \right)}\); this is equivalent to the ratio of the number of obfuscating \textit{apps} in $S_o$ (that need to be installed and used to protect privacy) and the size of original \textit{apps} set $S_a$. \textit{Cost} is therefore defined as:

\begin{equation}\label{formula:cost}
C={\left| {S_o} \right| } / {\left| {S_a} \right| }
\end{equation}

We consider the \textit{resource overhead} $R$, to be the overall resource usage by running the recommended \textit{apps}. Hence, there will be a corresponding overhead $R_{i,j}^t$ for each \textit{app} $a_{i,j}$ at time slot $t$, comprising broadly of \textit{communication} \(R_c^t\left( {{a_{i,j}}} \right)\), \textit{processing} \(R_p^t\left( {{a_{i,j}}} \right)\), and \textit{battery} consumption overheads \(R_b^t\left( {{a_{i,j}}} \right)\).

\begin{equation}\label{formula:resource-usage}
R_{i,j}^t = R_c^t\left( {{a_{i,j}}} \right) + R_p^t\left( {{a_{i,j}}} \right) + R_b^t\left( {{a_{i,j}}} \right)
\end{equation}

Following, we present further detail on resource usage and then experimentally evaluate various components (see Section \ref{sec:resouce-use-insights}) of the overall resource overhead $R$.

\subsection{Evaluating resource use}\label{sec:resouce-use-exp}
We rely on various utilities of \texttt{Android SDK}\footnote{\url{https://developer.android.com/studio}} to automate various measurements for \textit{processing} and \textit{battery} consumption. As an example, we execute \texttt{adb shell top -m 10} within the \sloppy \texttt{Process p = Runtime.getRuntime().exec("command")} to determine \textit{CPU} processing of each running \textit{app}. Similarly, to evaluate \textit{battery}'s current status, we use \texttt{adb shell dumpsys battery | grep level} command by initiating the \texttt{startActivity()} of the \texttt{Intent} utility of \texttt{Android SDK}. To measure this, we first charge the battery to 100\% and then run the \textit{app} for one hour, while connected to WiFi network. We envisage that running of recommended \textit{apps} would be used over WiFi to reduce communication costs (specifically for users with limited mobile network packages), although, in real life scenario, users are likely to utilise \textit{apps} (equally) on a mobile network that would results different magnitudes of resource overheads.

Similarly, we utilise the traffic captured during our experiments to evaluate \textit{communication} overhead; described in Section \ref{sec:experimentalSetup}.

In addition to above, we evaluate another resource usage overhead i.e. \textit{storage space} consumption. This can be further classified into \textit{installation} storage space, \textit{cache} storage i.e. temporary stored data, e.g. cookies stored on phone, internal \textit{data} size i.e. storage used for \textit{apps}' files, accounts, etc., which are respectively calculated using \texttt{codeSize}, \texttt{cacheSize}, and \texttt{dataSize} of \texttt{PackageStats} package of \texttt{Android SDK}. Note that we automated all these through an \textit{app}, and all the experimentation discussed in next section, using Android Debug Bridge\footnote{\url{https://developer.android.com/studio/command-line/adb}} of \texttt{Android SDK} that enables communication between a PC and connected \texttt{Android} devices.

\section{Performance Evaluation} \label{evaluation}

We now discuss details of various components of `System App' implementation and experimental evaluation and further provide insights on resource overheads.

\subsection{System App: The obfuscating system}\label{sec:obfusSystem}
We have implemented a POC `System App' of our proposed framework. Various components are presented on left side of Figure \ref{system-model} i.e. `User Environment', which introduces changes to the user side with a range of different functionalities e.g. it implements local user profiling both for \textit{Context profile} and \textit{Interests profiles} (as detailed in Section \ref{protect-sensitive-attr}), protects user's privacy for their (private) sensitive attributes, and preserves user privacy for their \textit{apps usage} behavior (Section \ref{prtect-apps-usage}). In addition, it implements our proposed online control algorithm for jointly optimising user \textit{privacy} and \textit{cost}; presented in Section \ref{online-control-algo}. To enable this functionality, a user need to install and run the `System App'; this approach is similar to existing \textit{app} recommender systems, e.g. AppBrain\footnote{\url{https://www.appbrain.com/}}. The `System App' acquires information about the set of currently installed \textit{apps} on a device, interacts with the user in regards to the selection of private attributes to be protected. Furthermore, it evaluates the list of candidates obfuscating \textit{apps} and presents to the user and automates the process of installation and running of these \textit{apps}. 

The `System App' sends the installed \textit{app} information to the `System Engine' for calculating obfuscating \textit{apps} and automates the installation and running of these \textit{apps}, as shown in Figure \ref{system-model}. We have used various utilities of the \texttt{Android SDK} in our implementation. E.g., \texttt{PackageManager}\footnote{\url{https://developer.android.com/}} is used to retrieve the list of installed \textit{apps}. The meta data of the installed \textit{apps}, such as \textit{app} name, permissions etc., are obtained by calling the function \sloppy \texttt{getInstalledApplications (PackageManager.GET\_META\_DATA)}. The `System Engine' evaluates obfuscating \textit{apps} according to the criteria discussed in Section \ref{protect-sensitive-attr} by examining the `Local \textit{app} repository'; we suggest that this repository is updated by the advertising system so that the `System App' can calculate obfuscating \textit{apps} from various \textit{app}'s categories.

The `System Engine' module forwards the list of obfuscating \textit{apps} to the `System App'. Each \textit{app} is displayed to the user as an accessible hyperlink to Google Play (or `Apple's App Store') store, which is done by invoking the \sloppy \texttt{startActivity (new Intent(Intent.ACTION\_VIEW, Uri.parse ("market://details?id=" +appPackageName)))}. The activity name (the \textit{app} name that can be recognised by app market of the \textit{app} to be installed) is specified using \texttt{appPackageName} function using the \texttt{Intent} class. These \textit{apps} are automated and run using the \texttt{startActivity} utility of \texttt{Android SDK} for a specified amount of time that is required to generate new profile interests in the ad system.

\subsection{Experimental setup}\label{sec:experimentalSetup}
In this work, we mainly focus on Google AdMob\footnote{\url{https://apps.admob.com}} since it is the leading marketplace in mobile user profiling and has captured the online digital advertising market, however, we note that the proposed methods can be effortlessly applied to other ad/analytic networks. We carry out various experimentations for preserving profiling privacy (i.e. \({I_g^t + \eta _{l'}^t\left( {{\Psi _{l'}}} \right)}\)), \textit{apps usage} (\({I_g^t - U{'^t}\left( {{K_a}} \right)}\)) and evaluating various resource overheads $R$. To evaluate this, we select \textit{apps} from 27 random (sample) categories; we note that Google Play store has organised \textit{apps} into 37 (considering `Games' as a single category) various categories e.g. `Entertainment', `Lifestyle' etc. For these experimentations, we select top 100 free \textit{apps} from randomly chosen 27 categories and further narrow down it to 10 highest ranked \textit{apps} from each category. As mentioned earlier that the Google user profile is partly based on mobile \textit{apps} usage and relevance of received ads, hence, we had to ensure that the tested \textit{apps} receive ads.

We set a second (\textit{mapping}) experimental setup; we select one mobile \textit{app} from the top list of any category and run it for a period of up to 96 hours; note that this process was automated, as described above in Section \ref{sec:obfusSystem}. The purpose of these experiments was to evaluate \textit{mapping} of specific \textit{Context profile} to \textit{Interest profile} i.e. \({K_a} \to {I_g}\), as discussed in Section \ref{representing-user-profile}. This helps in determining contribution of individual \textit{app} in an \textit{Interest profile}. We note that findings from these experiments help in selecting recommended \textit{apps} for various disturbances in user profiles i.e. achieving different trade-offs between \textit{privacy} and \textit{targeted ads}, as explained in Eq. (\ref{first-scenario}), (\ref{third-scenario}), and (\ref{obfuscation-weightage}) e.g. $0 < {\eta _{l'}}\left( {{\Psi _{l'}}} \right) \le \eta _l^{\min }\left( {{\Psi _l}} \right)$ for lower \textit{privacy} and higher \textit{targeted ads}. These experiments have taken around 3 months to complete for the 270 highest ranked \textit{apps} from 27 random \textit{apps} categories.

The ads traffic, including the control traffic, exchanged for tracking/profiling purposes, was collected using \texttt{tcpdump}\footnote{\url{www.tcpdump.org}}, cleansed and saved to a local database during entire experimentations. We reset the profile\footnote{The profile was reset by using `Reset advertising ID' option in \textit{Google Settings} system app i.e. \texttt{Google Settings} $\to$ \texttt{Ads} $\to$ \texttt{Reset advertising ID}.} before starting each experiment in order to make sure that the \textit{Interest profile} is only resulting from the currently installed and actively used \textit{apps}. In addition, we set up a phone with the same configuration however with `Opt-out of Ads Personalisation' enabled in \textit{Google Settings} system app. The purpose of this phone is to have a base reference for both newly generated user profiles and received \textit{targeted ads}. These experiments were run for all the selected categories 24/7 for 5 months; due to the practical limitations of these experiments, we only use 10 smartphones in parallel.

We also used the collected traffic from these experiments for calculating the \textit{resource usage}; detailed in Section \ref{sec:resouce-use-exp}.

\begin{figure*}[h]
   \begin{center}
   \subfigure[Lower profile disruption]{
      \includegraphics[width=0.82\columnwidth]{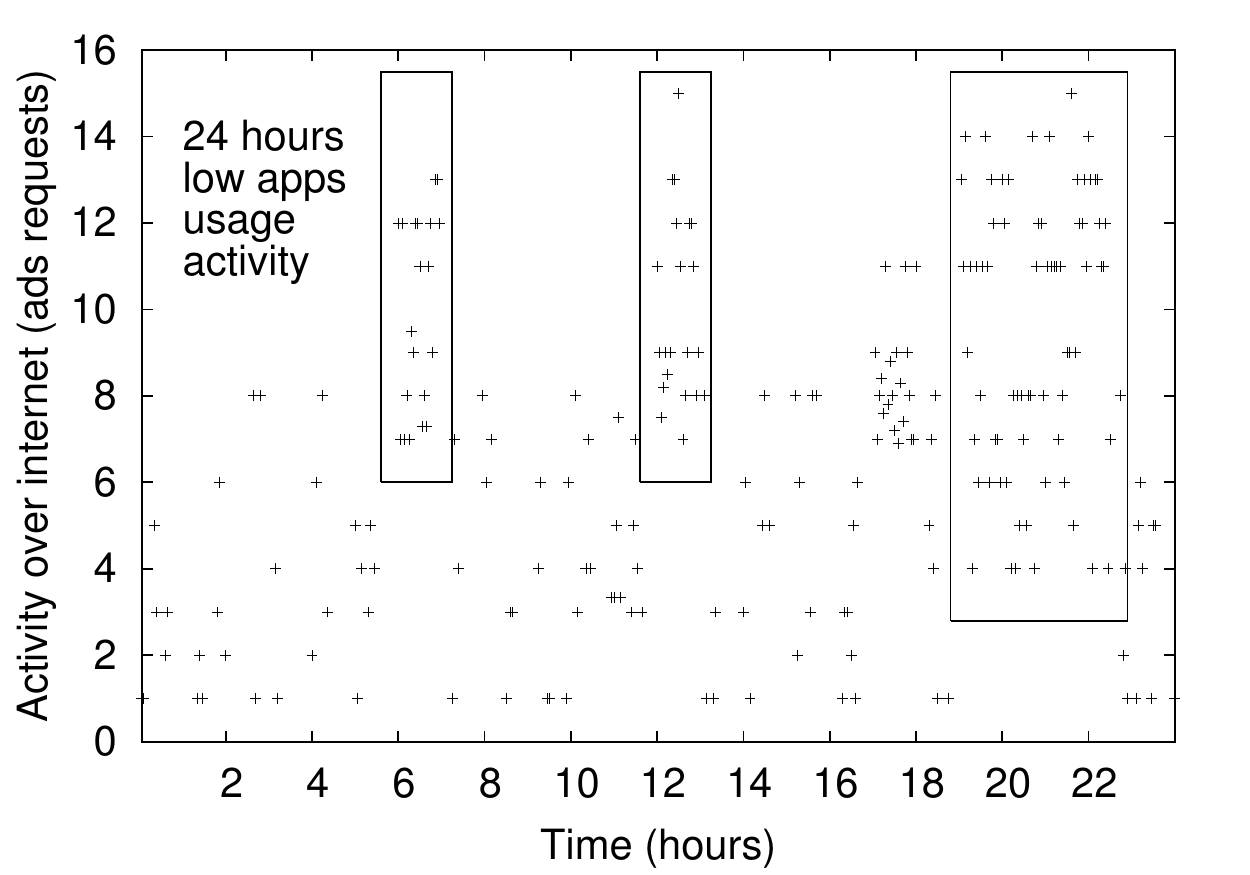}
    }
    \subfigure[Higher profile disruption]{
 			\includegraphics[width=0.82\columnwidth]{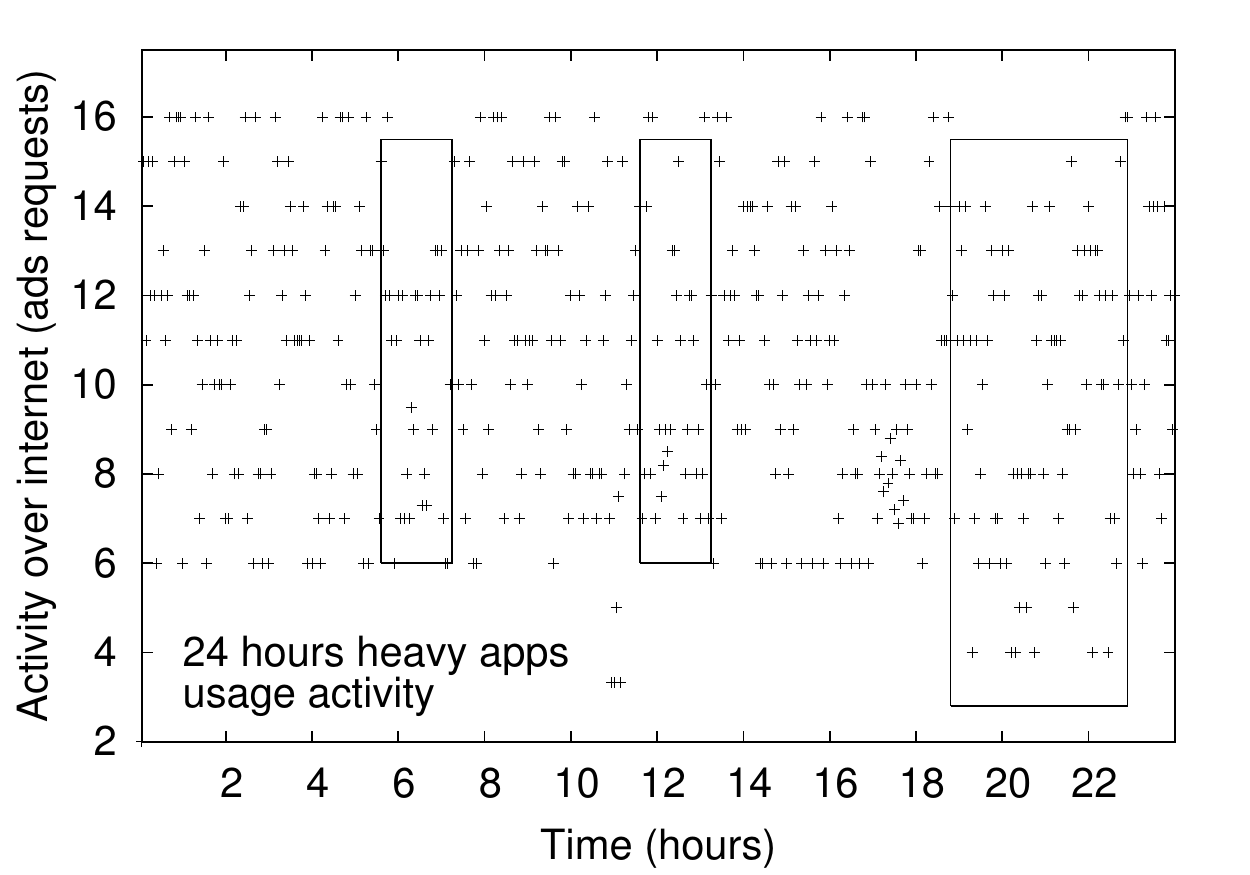}
    }
   \end{center}
  \caption{Lower activity of recommended \textit{apps} for lower (a) and higher (b) profile disruption.}
   \label{fig:profile-disruption}
\end{figure*}

\subsection{Trade-off between apps usage privacy and cost of profile disruption}\label{sec:profile-disruption}
Figure \ref{fig:profile-disruption} compares the \textit{apps} usage profile for lower and higher profile disruption by introducing lower (as shown in (a)) and higher (i.e. (b)) activity of recommended \textit{apps}; note the original user profile activity is shown with rectangular boxes; already discussed in Figure \ref{apps-usage-behavior}. We first divide the time into 5 min duration bins and then record the number of ads requests in each bin (there are different number of ad requests by each \textit{app}; as detailed in Section \ref{sec:resouce-use-insights}); each dot (in plot) in both Figures \ref{fig:profile-disruption} (a) and (b) show the ad request frequency during the 24 hours time (shown on x-axis) and their corresponding ads requests frequency (i.e. shown on y-axis).

The recommended \textit{apps} were run during various (day/night) times where there was no original ($K_a$) \textit{app}'s activity e.g., during 1am--6am, as indicated in both figures. We note that, during 1am--6am with `lower \textit{app} usage and low profile disruption', newly \textit{apps} request 27 ad (binned) requests with a total of fetching 93 ads, while it is respectively 93 and 1323 for `higher \textit{app} usage and high profile disruption'. In addition, the average number of ad requests and corresponding ads fetched were 19.6 and 60 for `lower \textit{app} usage and low profile disruption', while it was 77.6 and 809.6 for `higher \textit{app} usage and high profile disruption'. Overall, there were 340 and 5068 ad requests respectively with two options of profile disruptions; note that this also has an effect over various overheads, in particular, the \textit{Communication} overhead, also discussed later in Section \ref{sec:resouce-use-insights}.

\subsection{Privacy protection vs. ads distribution}\label{sec:apps-usage-privacy}
\begin{figure*}[h]
   \begin{center}
   \subfigure[Lower \textit{apps} usage and lower ads distribution]{
      \includegraphics[width=0.8\columnwidth]{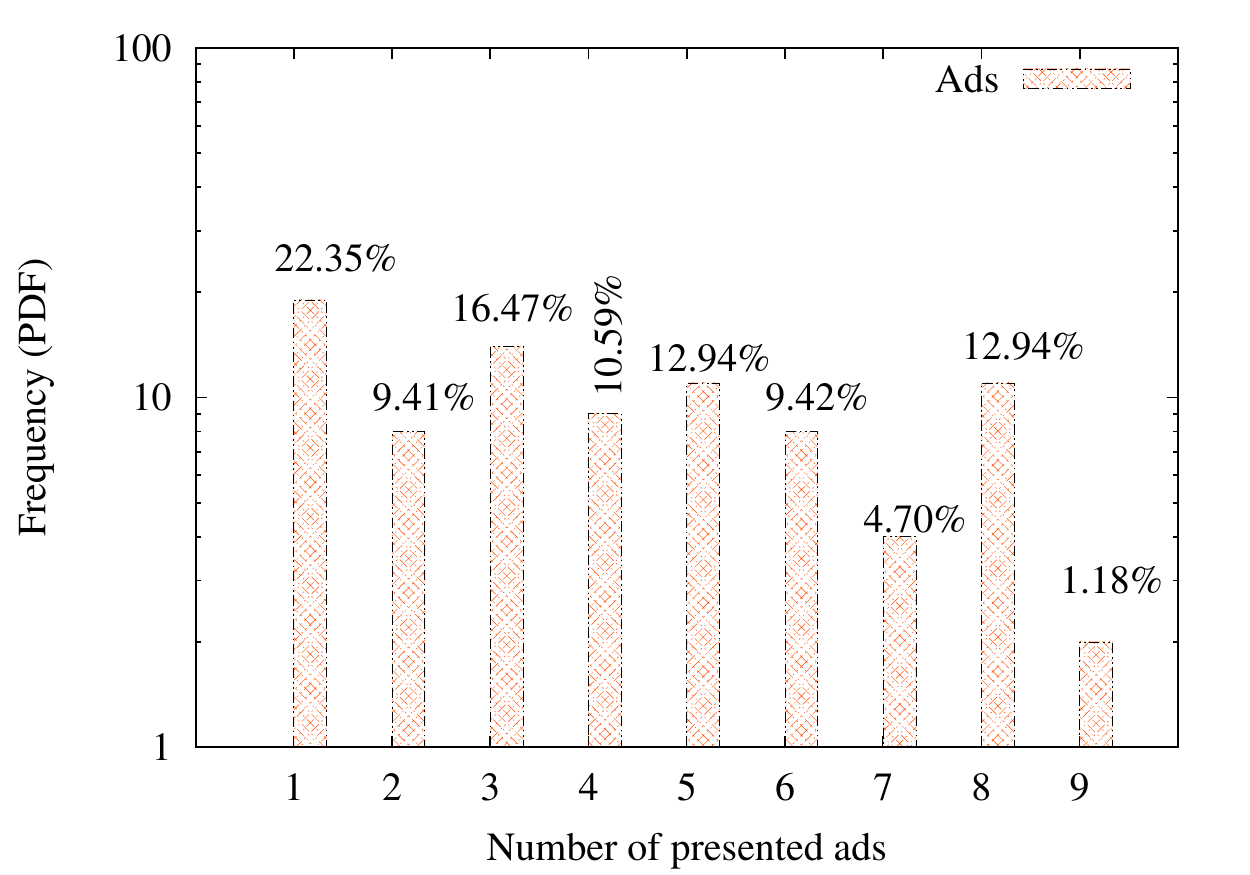}
    }
    \subfigure[Higher \textit{apps} usage and higher ads distribution]{
 			\includegraphics[width=0.8\columnwidth]{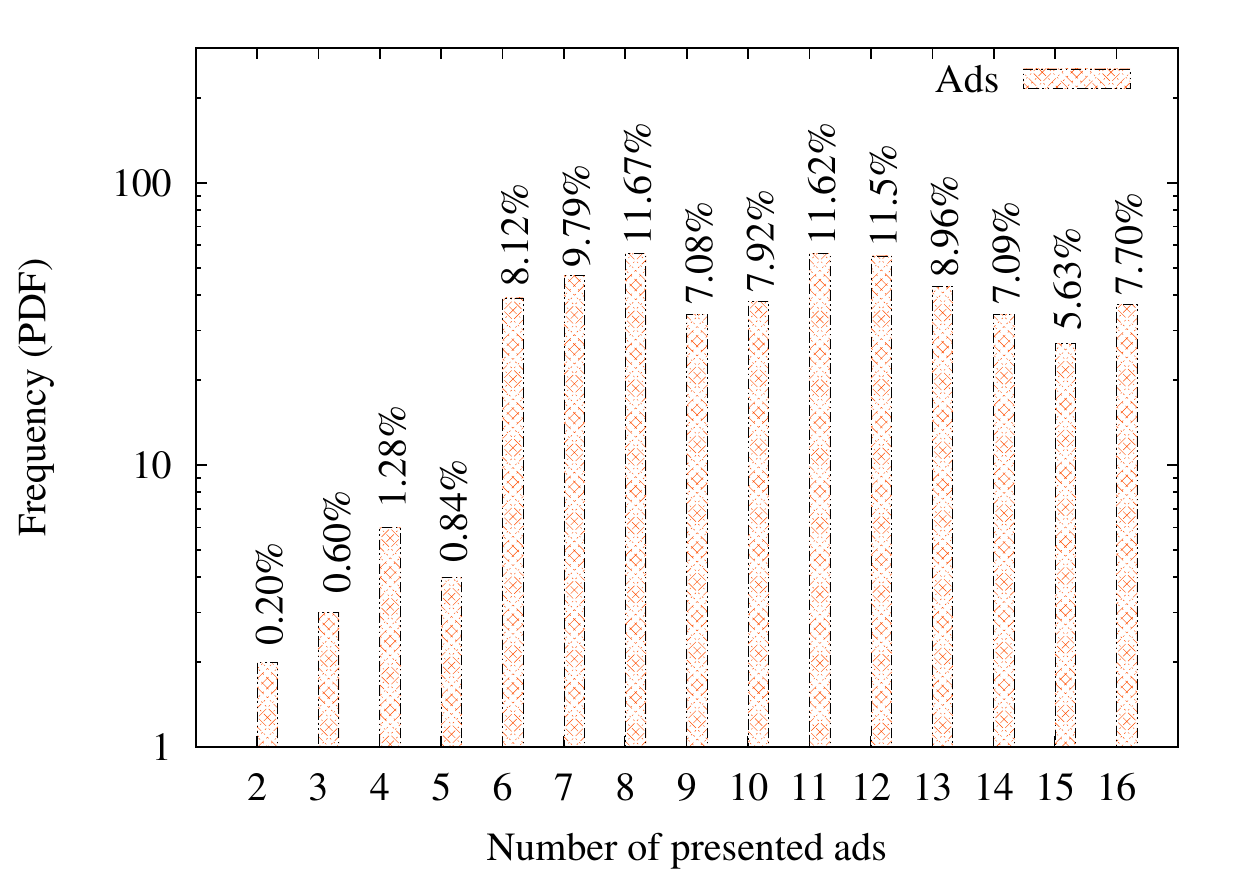}
    }
   \end{center}
  \caption{Distribution of ads by lower (a) and higher (b) \textit{apps} usage.}
   \label{fig:apps-usage-privacy}
\end{figure*}

Figure \ref{fig:apps-usage-privacy} (a) and (b) show frequency distribution of ads for lower and higher \textit{apps} usage; the percentage of ads for presented ads in different time-bins is also shown on top of each bar. For each time bin, the presented ads were ranged from 1 to 16 ads requests, in particular, the frequency of these requests range from 1 to 9 ads requests for `lower \textit{apps} usage', which also attracted a lower number of ads i.e. only 85 ads to completely \textit{apps} usage privacy as shown in Figure \ref{fig:apps-usage-privacy} (a). Similarly, for `higher \textit{apps} usage', the ad requests range from 1 to 16 per time bin with an average of 32 ads for all frequency of ads, with a total of 480 ads to completely preserve user's privacy. Recall that, the selection of such \textit{apps} usages are mainly done for two purposes i.e. to protect \textit{apps usage privacy} and to affect the \textit{targeted ads}.

\subsection{Privacy protection vs. resource use}\label{sec:resouce-use-insights}
It is reported by previous works \cite{vallina2012breaking, ullah2014profileguard}, and is also evident in our experimental results, that the ads (and its related tracking traffic) are the major contributor to the \textit{communication} \(R_c^t\left( {{a_{i,j}}} \right)\) overhead. Hence, we approximate the \textit{communication} cost by the traffic generated by ads and their related actions by users.

We note that, from the bandwidth viewpoint, the ads traffic is characterised by various components: The \textit{ad refresh rate}\footnote{AdMob refers this to \textit{rate}, which is deterministic for every \textit{app} and is configured by \textit{apps} developer at the time of registering \textit{apps} on Google Play store.} (technically it is the inter-arrival time of two consecutive ads); their correspondence with various ad/analytic servers with an advertising ecosystem; contacts with \texttt{CDN} for downloading various ad components e.g. images etc.; the number of objects associated with an ad along with their sizes; and communication with various servers during interactions with an ad. Table \ref{table:reqresponses} shows various ad-related objects and control messages, along with their sizes; determined from collected traffic traces.

During our experimentations, we examined, from collected traces, that an ad size is \(16 \pm 4\)KBs, which (on average) contains 8--10 objects (e.g. \texttt{JavaScript} files, images, etc.) with an average of 30--35 request/response messages. In addition, we note that \textit{ad refresh rates} vary between 20--60 seconds\footnote{We note that supported values are 12-120 seconds in Google AdMob.}, with distinct values of 20, 30, 45 and 60 seconds, which respectively to 36\%, 47\%, 15\% and 2\% of all the tested \textit{apps}. Since the ad sizes do not vary widely, hence the \textit{communication} overhead for introducing lower disturbance in a user profile can be further minimised by selecting those \textit{apps} that have maximum overall \textit{ad refresh rate}; note that this information is already available with any ad network e.g. Google AdMob.

\begin{table}[h]
\caption{The ad-related objects and control messages along with their average size in bytes.}
\begin{center} \scalebox{0.7}{
\label{table:reqresponses}
{\begin{tabular}{|l|c|l|c|}
\hline
\textbf{Message Type} & \textbf{Size (bytes)} & \textbf{Message Type} & \textbf{Size (bytes)}\tabularnewline
\hline
200 Ok & 496 & POST & 350\tabularnewline
\hline
DNS Query Request & 68 & Ok/(PNG) & 1300\tabularnewline
\hline
GET /pagead/images & 578 & Ok/(GIF) & 1000\tabularnewline
\hline
DNS Query Response & 334 & Ok/(JPEG) & 1300\tabularnewline
\hline
Ok/(text/html) & 1200 & TCP reassembled PDU & 1434\tabularnewline
\hline
GET /geocode & 224 & no content & 396\tabularnewline
\hline
GET /simgad & 252 & Ok/(application/json) & 240\tabularnewline
\hline
GET /mads/gma & 685 & Ok/(text/javascript) & 800\tabularnewline
\hline
GET /imp & 244 & Ok/(text/css) & 824\tabularnewline
\hline
GET /generate\_204 & 244 & TCP Ack & 66\tabularnewline
\hline
GET /csi & 595 & TCP Syn & 74\tabularnewline
\hline
\end{tabular}}}
 \end{center}
\end{table}

\begin{figure}[h]
\begin{center}
\includegraphics[scale=0.55]{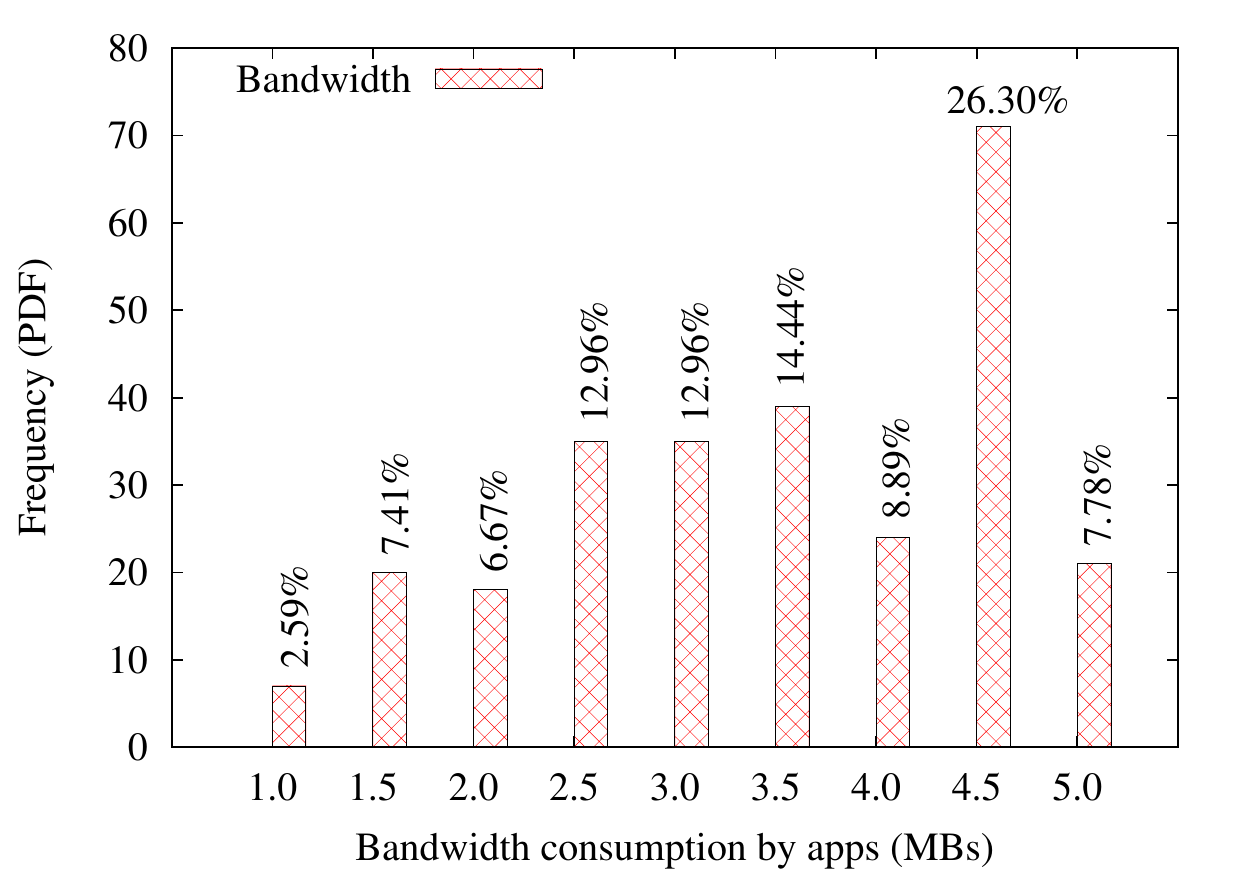}
\caption{Bandwidth consumption (MB) calculated for ads by tested \textit{apps} during experimentations.}
\label{fig:ads-bandwidth-optimsn-hist}
\end{center}
\end{figure}

Figure \ref{fig:ads-bandwidth-optimsn-hist} shows the distribution of bandwidth used by \textit{apps} during our experimentations per our experimental setup, shown in Section \ref{sec:experimentalSetup}. The proportion of \textit{apps} that have consumed the respective bandwidths are also shown on top of each bar e.g. 26.30\% of the \textit{apps} (which corresponds to 71 \textit{apps}) consumes a high bandwidth of 4.0--4.5MB. In addition, we note that \textit{apps} that frequently fetch ads i.e. \textit{apps} with \textit{ad refresh rates} of 20 and 30 seconds, utilise an average bandwidth of 3.0--5.5MB; these \textit{apps} represent around 70\% of all the experimented \textit{apps}. Note that the selection of such \textit{apps} can be used for introducing higher disruption in user profiles, to attract lower \textit{targeted ads}, and to achieve high \textit{apps usage privacy}. The remaining 30\% of the \textit{apps} (i.e. \textit{apps} with \textit{ad refresh rates} of 45 and 60 seconds) utilise between 0.5--2.5MB communication bandwidth.

Subsequently, we evaluate the \textit{processing} overhead introduced by experimented \textit{apps}. We note that the measured \textit{CPU usage} varies, although not widely, across different \textit{apps}: the CPU-intensive \textit{apps} such as those from the `Games' category use between 25\% to 30\% of the \textit{CPU} power; less-interactive \textit{apps} such as \textit{Notes} use between 15\% to 20\%.

Similarly, the \textit{battery} consumption measured in our experiments shows a relatively low variations between various \textit{apps}, with between 30\% to 40\% of the total battery (i.e. 100\%) being used by each \textit{app} during the measurement period.

In addition to above basic \textit{overheads}, we evaluated \textit{storage space} consumed by recommended \textit{apps}; we mainly determined the \textit{installation}, \textit{data}, \textit{cache} storage spaces. These storage spaces vary for \textit{apps} according to their requirements e.g. a `language translation' \textit{app} might do offline text translation, hence it would require to save library files in \textit{data} storage space quota requiring more space compared to \textit{installation} storage space. In contrast, the \textit{Facebook} would consume more \textit{data} storage space to store user accounts data, search history, group settings, user timeline data etc. In addition, a \textit{Google Maps} would take more \textit{cache} storage space to save user's searched places history etc. Table \ref{table:apps-storages} presents few representative \textit{apps} for combination of these storage space requirements; following we present further details.

\begin{table}[h]
\caption{Representative \textit{apps} with various combination of storage space requirements.}
\begin{center} \scalebox{0.6}{
\label{table:apps-storages}
{\begin{tabular}{|c|c|c|c|c|c|c|c|c|}
\cline{1-4} \cline{6-9}
\multirow{2}{*}{\textbf{Apps}} & \multicolumn{3}{c|}{\textbf{Storage space consumed}} & \multirow{9}{*}{} & \multirow{2}{*}{\textbf{Apps}} & \multicolumn{3}{c|}{\textbf{Storage space consumed}}\tabularnewline
\cline{2-4} \cline{7-9}
 & \textbf{Installation} & \textbf{Data} & \textbf{Cache} & & & \textbf{Installation} & \textbf{Data} & \textbf{Cache}\tabularnewline
\cline{1-4} \cline{6-9}
Youtube & 117 & 2.57 & 361 & & Subway Surf & 156 & 85.98 & 15.57\tabularnewline
\cline{1-4} \cline{6-9}
Chrome & 86.88 & 15.46 & 1.53 & & File Manager & 8.37 & 28.04 & 17.83\tabularnewline
\cline{1-4} \cline{6-9}
Foxit PDF & 96.42 & 1.17 & 32.91 & & London City Guide & 60.98 & 88.12 & 4.56\tabularnewline
\cline{1-4} \cline{6-9}
Google Play Store & 93.61 & 7.48 & 22.25 & & Skype & 62.95 & 28.22 & 13.41\tabularnewline
\cline{1-4} \cline{6-9}
Babel & 51.98 & 14.79 & 0.02 & & Viber & 159 & 17.35 & 0.12\tabularnewline
\cline{1-4} \cline{6-9}
Google Translate & 5.53 & 392 & 2.04 & & Adobe Acrobat & 20.91 & 0.47 & 0.22\tabularnewline
\cline{1-4} \cline{6-9}
Amazon Kindle & 53.30 & 123.33 & 4.98 & & TripView Lite & 28.21 & 5.92 & 2.09\tabularnewline
\cline{1-4} \cline{6-9}
\end{tabular}}}
 \end{center}
\end{table}

\begin{figure*}[h]
   \begin{center}
   \subfigure[Installation Storage]{
      \includegraphics[width=0.65\columnwidth]{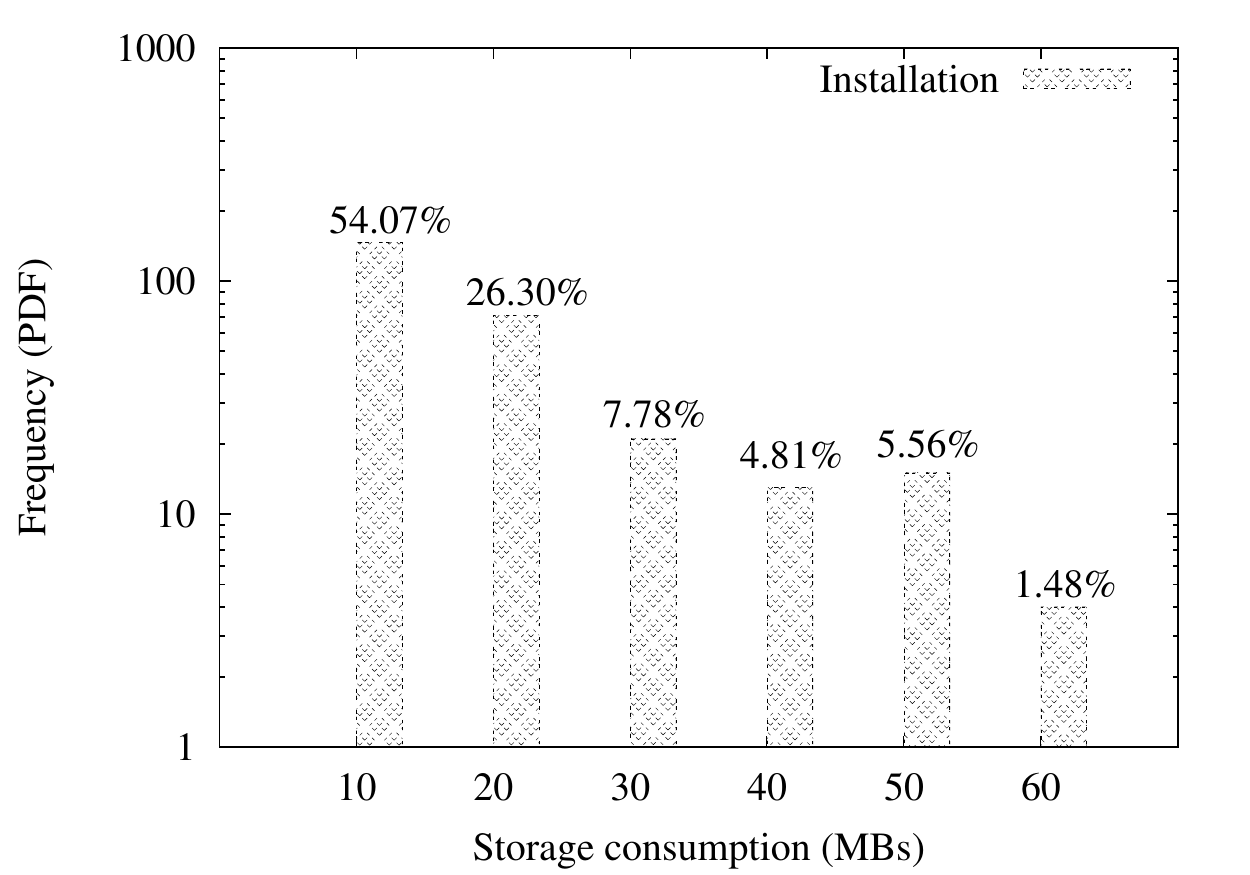}
    }
    \subfigure[Data Storage]{
 			\includegraphics[width=0.65\columnwidth]{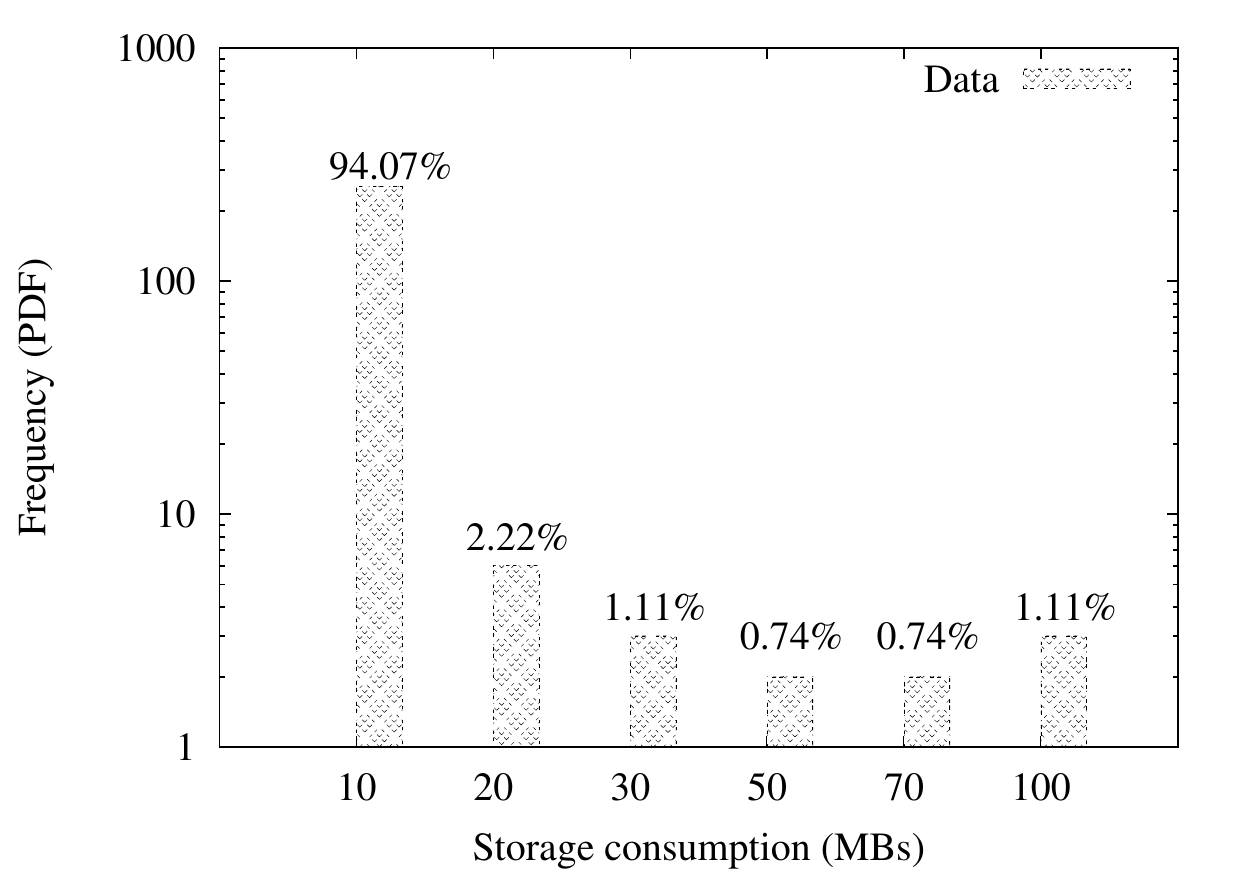}
    }
   \subfigure[Cache Storage]{
      \includegraphics[width=0.65\columnwidth]{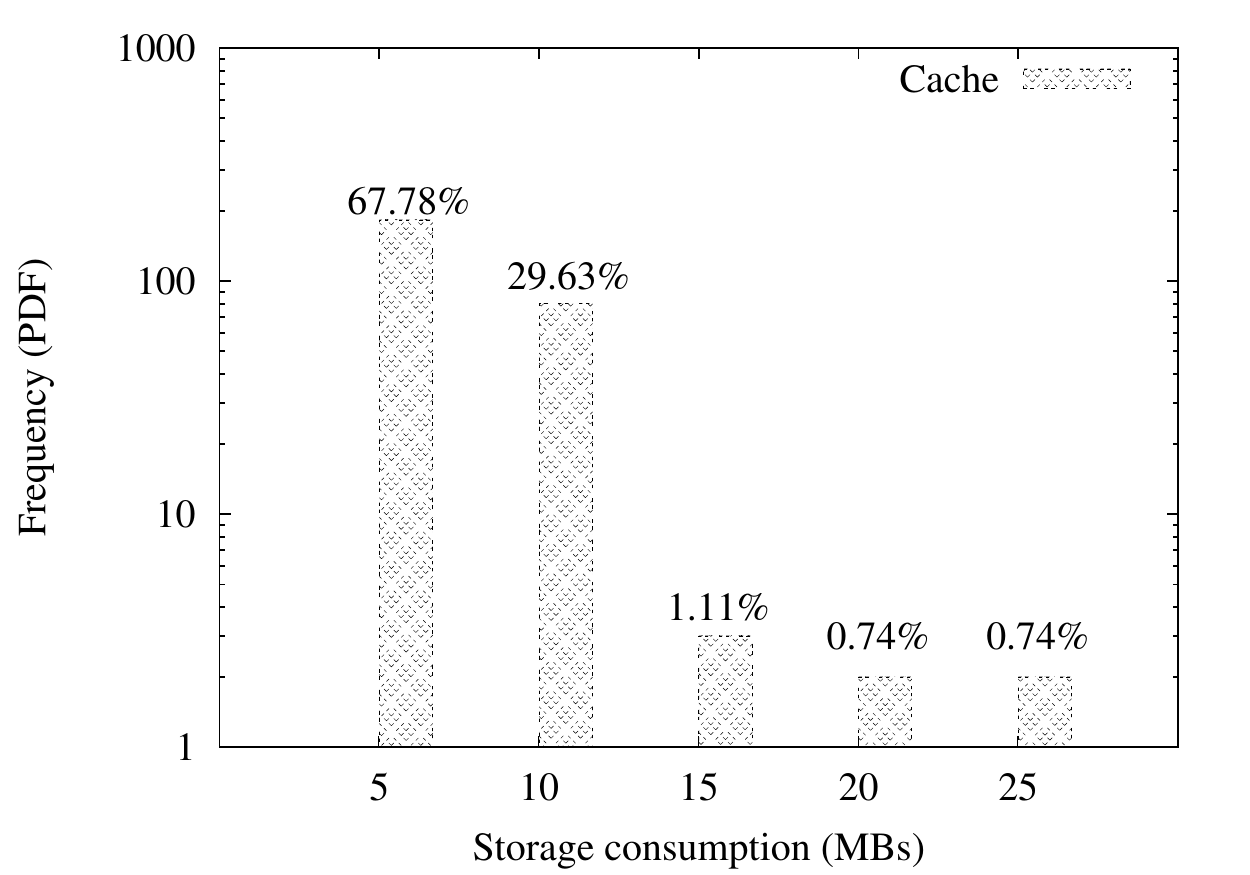}
    }
   \end{center}
  \caption{Storage space consumed by experimented \textit{apps} for (a) \textit{Installation} (b) \textit{Data} and (c) \textit{Cache}.}
   \label{fig:ads-storage-consumptions-hist}
\end{figure*}

Figures \ref{fig:ads-storage-consumptions-hist}(a) through (c) show the distribution of \textit{storage spaces} of the experimented \textit{apps}; the proportion of \textit{apps} that requires respective storage space are also shown on top of each bar. It can be observed, for \textit{installation} storage, that nearly half (54\%) of the experimented \textit{apps} acquires lower \textit{storage space} i.e. 0.5MB to 10MB, while only 1.48\% of these \textit{apps} consume relatively higher \textit{storage space} of 50--60MB. Similar observations can be found for \textit{data} and \textit{cache} overhead, as shown in Figures \ref{fig:ads-storage-consumptions-hist} (b) and (c). In conclusion, we observe that vast majority of these \textit{apps} use a lower amount of \textit{storage space}; e.g., 80\%, 97\%, and 98\% of all \textit{apps} belong to the lowest \textit{storage space} bins, i.e. within the range of 0.5--20MB, 0.01--20MB, and 0.01--10MB storage consumption, respectively for \textit{installation}, \textit{data}, and \textit{cache} storage overheads.

\section{Discussion} \label{discussion}

We discuss the applicability of our framework in the current advertising ecosystem, in addition to, comparing the proposed framework for various privacy protection approaches for the presented threat model.

\subsection{Protecting sensitive profiling interests via Differential privacy}\label{differntial-privacy}
The concept of \textit{differential privacy}\footnote{A C++ implementation of \textit{differential privacy} library can be found at \url{https://github.com/google/differential-privacy}.} was introduced in \cite{dwork2006calibrating}, a mathematical definition for privacy loss associated with any released data fetched from a database. A deeper understanding of \textit{differential privacy} and its algorithms can be found in \cite{dwork2014algorithmic}. Let \({D_1} = {g_{k,l}} \in S\) i.e. the set of interests in a user profile, and \({D_2} = g{'_{k,l}} \in {S_g}\), where \({\Psi _l} \ne \Psi {'_l}\) and \(g{'_{k,l}}\) is the profiling interest(s) other than the primary set of interests defined by an advertising company, \({\Psi _l}\) is a selected private interests category that the user wants to protect. Subsequently, the \textit{randomised} function $K$ gives \textit{differential privacy} for these two data sets as:

\begin{equation}\label{eq:differntial-privacy}
{{\mathop{\rm P}\nolimits} _r}\left[ {K\left( {{D_1}} \right) \in S} \right] \le \exp \left( \varepsilon \right) \times {{\mathop{\rm P}\nolimits} _r}\left[ {K\left( {{D_2}} \right) \in S} \right]
\end{equation}

We examine the use of other privacy protection mechanisms that can also be utilised e.g. \textit{private information retrieval}, \textit{anonymisation}, \textit{randomisation}, \textit{Blockchain-based} solutions, however, we note that these solutions do not fully protect user's privacy for the presented threat model; following we provide a comparative analysis of the use of various other privacy protection mechanisms.

\subsection{Comparison with other privacy protection approaches}\label{applicability-comparison}
Table \ref{privacy-comparision} provides a hypothetical comparison of various privacy protection mechanisms using different parameters, evaluated in our proposed framework. It can be observed that only the proposed mechanism of introducing recommended obfuscation \textit{apps} protects the user's privacy for `app's usage behavior' since the user has to run these apps in order to protect usage behavior at different periods of the day/night. Similarly, an important parameter is the `trade-off between privacy and targeted ads', which can only be achieved using the proposed mechanism and the \textit{randomisation} and \textit{obfuscation}. Furthermore, another parameter is to protect `user privacy in terms of serving targeted ads' (an \textit{indirect} privacy attack to expose user privacy), which can be adjusted according to user's needs i.e. `low-relevant vs. high-relevant interest-based ads'. We plan to carry out a comprehensive study over these parameters for these various privacy protection mechanisms in the future in order to validate/invalidate our hypotheses.

\begin{table*}[h]
\begin{center} \scalebox{0.90}{
{
\begin{tabular}{|l|c|c|c|c|c|c|c|}
\hline
\multirow{2}{*}{\textbf{Parameters}} & \multirow{2}{*}{\textbf{Proposed}} & \textbf{Differential} & \textbf{Cryptographic} & \multirow{2}{*}{\textbf{Randomisation}} & \textbf{Obfuscation} & \textbf{Blockchain-based} & \multirow{2}{*}{\textbf{Anonymisation}}\tabularnewline
 & & \textbf{Privacy} & \textbf{Mechanisms} & & \textbf{(Profiling)} & \textbf{Solutions} & \tabularnewline
\hline
\hline
Apps usage behaviour & \multirow{2}{*}{Guaranteed} & \multirow{2}{*}{No guarantee} & \multirow{2}{*}{No guarantee} & \multirow{2}{*}{No guarantee} & \multirow{2}{*}{No guarantee} & \multirow{2}{*}{No guarantee} & \multirow{2}{*}{No guarantee}\tabularnewline
privacy & & & & & & & \tabularnewline
\hline
\multirow{2}{*}{Profiling privacy} & Yes & \multirow{2}{*}{Yes (Low)} & \multirow{2}{*}{Yes} & \multirow{2}{*}{Yes (Low)} & \multirow{2}{*}{Yes (Low)} & \multirow{2}{*}{Yes} & \multirow{2}{*}{Yes}\tabularnewline
 & (Low to high) & & & & & & \tabularnewline
\hline
Indirect privacy exposure & Yes & \multirow{2}{*}{Yes} & \multirow{2}{*}{No} & Yes & \multirow{2}{*}{Yes} & \multirow{2}{*}{No} & \multirow{2}{*}{Yes}\tabularnewline
from targeted ads & (Low to high) & & & (Low to high) & & & \tabularnewline
\hline
Cost \$C\textasciicircum{}t\$ & Low to High & Low & High & High & High & High & High\tabularnewline
\hline
\multirow{2}{*}{Targeted ads} & Minor to no relevant & \multirow{2}{*}{Low} & \multirow{2}{*}{Yes} & \multirow{2}{*}{Low} & \multirow{2}{*}{Low} & \multirow{2}{*}{Yes} & \multirow{2}{*}{Yes}\tabularnewline
 & ads (adjustable) & & & & & & \tabularnewline
\hline
Tradeoff b/w privacy and & \multirow{2}{*}{Yes} & \multirow{2}{*}{Yes} & \multirow{2}{*}{No} & \multirow{2}{*}{Yes} & \multirow{2}{*}{Yes} & \multirow{2}{*}{No} & \multirow{2}{*}{No}\tabularnewline
targeted ads & & & & & & & \tabularnewline
\hline
Impact over billing for & Yes & Yes & \multirow{2}{*}{No} & Yes & Yes & \multirow{2}{*}{No} & \multirow{2}{*}{No}\tabularnewline
targeted ads & Low to High & Low & & Low to Hig & Low to Hig & & \tabularnewline
\hline
\end{tabular}
}}
 \end{center} \caption{Comparison of proposed framework with other privacy protection mechanisms for various important parameters applicable in advertising ecosystems. \label{privacy-comparision}}
\end{table*}

\subsection{Applicability of proposed framework in advertising system}\label{applicability}
\subsubsection{\textbf{Overall System Functionality}}
The motivation for protecting user privacy is very much dependent on the way consumers use mobile \textit{apps} and access the internet; users are ever more concerned about preserving their privacy due to an enormous increase in its awareness. Such an example of awareness motivation is by exposure of mass surveillance activities \footnote{\url{http://www.theguardian.com/world/the-nsa-files}} and by unauthorised leaks of personal data. Hence, users have ever more interested in the use of personal (bespoke) privacy tools. Thus the proposed framework is considered, inline with the current \textit{apps} recommender and personalisation systems, that not only suggest usably recommended \textit{apps} but also enable an optimised and cost-effective privacy protection mechanisms. As we have seen in Section \ref{sec:profile-disruption}, this framework optimises the cost by selecting various number of \textit{apps} for protecting \textit{apps} usage privacy, protects user's private interest profile and achieves optimal trade-off between privacy protect and the \textit{targeted ads}. Furthermore, this framework ensures that the recommended \textit{apps} have an overall good \textit{usability} to users, based on similarity metric discussed in Section \ref{sec:privacyMetric}.

An important user's concern is the issue related to \textit{resource} use, we note during our \textit{mapping} experimentation (as detailed in Section \ref{sec:experimentalSetup}) that these experiments can help to significantly reduce the \textit{resource} use by selecting appropriate \textit{apps} that could (subject to availability of such \textit{mapping} information) effectively preserve user privacy with least overhead. Using these experiments, on average, the \textit{communication} cost is around 3MB (for lower profile disruption) compared to 17MB for higher profile disruption, to achieve \textit{apps} usage and profiling privacy. This would motivate users to use such a strategy, in particular, those who might be on a fixed-mobile data plan. As discussed earlier that AdMob profiling is based on `Web \& App' activity, hence such (\textit{mapping}) information can also be added for user's web searches/histories; although this \textit{app}-based strategy is still applicable to protect user privacy for interest profiling as profiling is now done via both `Web \& App' activity. We envisage that such information could be made derived by approximating user profile based on various related information available, such as search/history keywords, similar to interest \textit{mapping} discussed in Section \ref{representing-user-profile} for $K_a$.

\subsubsection{\textbf{Server Side Modifications}} Integrating the proposed framework within the existing advertising ecosystem would be fairly straightforward and would only require upgrading tracking and analytics mechanism of the server-side i.e. we suggest the advertising system transfer such functionalities to the client side; where the client remains honest. These changes are mainly related to \texttt{Aggregation} and \texttt{Analytics} servers; these servers would respectively receive the constructed user profile (along with anonymous \textit{apps} usage statistics without including user's Advertising ID) and other statistics, e.g. ad impression/clicks, etc., required by both advertising system and \textit{apps} developers.

\subsubsection{\textbf{Client Side Modifications}} A major change on the client side would require implementation via `System App' i.e. user \textit{profiling}, optimising user's privacy for \textit{targeted ads}, their \textit{interactions with ads} and \textit{app usage privacy} including the `Obfuscation engine' for selection of recommended \textit{apps}. Currently, the mobile integrates with the advertising ecosystem via \texttt{SDK}, hence it will mainly require modifications in client's AdMob \texttt{SDK}.

\subsection{\textbf{\hl{Research Limitations}}} \hl{The proposed system protects users' privacy against legitimate user profiling and the traffic monitoring/analysis and network surveillance both form the host advertising systems and third-party tracker/analytics. However, we did not address location-based privacy for \textit{ads \, targeting} in much more detail, although this can also be handled by our privacy protection approach since the location is added as one of the interests in user profile as part of the demographics, as detailed in Section} \ref{user-profiling}. \hl{Henceforth, for location-specific ads, low-resolution GPS coordinates can be included in user profile to accommodate advertisers and businesses wishing to advertise to passing trades. In addition, location can be protected using other protection mechanisms such as `Tor'; to use in conjunction with our system to prevent such threats.}

\section{Conclusion} \label{conclusion}

The online mobile \textit{targeted advertising} is growing in popularity, which at the same time has raised serious privacy concerns among individuals, enabled via the use of excellent user-tracking tools. This paper presents an optimal privacy-preserving and cost-effective framework for preserving user privacy due to user profiling, ads-based inferencing, and \textit{ads targeting} and (in general) user's behavior over mobile devices. We present a dynamic optimisation framework by first examining the underlying advertising ecosystem and then providing a privacy-preserving framework for temporal changes that occur in the user environment in an ad ecosystem. The online control algorithm is used to detect temporal changes in user profiling based on \textit{Lyapunov optimization} to achieve an optimal solution without requiring any knowledge of future use of mobile \textit{apps} or temporal changes in a user profile. We carry out extensive experimentations using mobile devices of various profiles and we examine the profiling process, privacy leakage, privacy protection, and resource usage. We develop a POC `System App' that implements critical components of the proposed framework and further discuss its applicability in an online advertising ecosystem.

\begin{IEEEbiography}[{\includegraphics[width=1in,height=1.25in,clip,keepaspectratio]{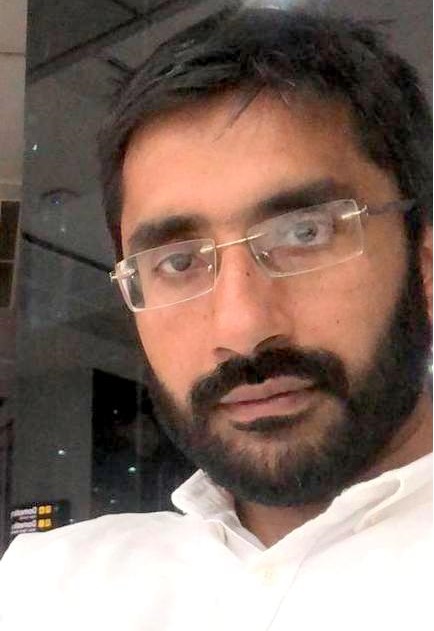}}]{Imdad Ullah} received his Ph.D. degree in Computer Science and Engineering from the University of New South Wales (UNSW), Sydney, Australia. He has served as on various research positions at UNSW, a Research Scholar at National ICT Australia (NICTA), Data61 CSIRO Australia, NUST, Islamabad, Pakistan, and SEEMOO TU Darmstadt, Germany, and a Research Collaborator at the SLAC National Accelerator Laboratory, Stanford University, USA. He is currently an Assistant Professor with the College of Computer Engineering and Sciences, PSAU, Saudi Arabia. He has research and development experience in privacy preserving systems, including private advertising and crypto-based billing systems. His primary research interest includes privacy enhancing technologies; he also has interest in the Internet of Things, Blockchain, network modeling and design, network measurements, and trusted networking.
\end{IEEEbiography}

\begin{IEEEbiography}[{\includegraphics[width=1in,height=1.25in,clip,keepaspectratio]{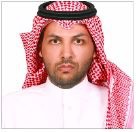}}]{Adel Binbusayyis} is currently an Assistant Professor in Computer Science at Prince Sattam Bin Abdulaziz University. He received his PhD degree from the University of Manchester, UK in 2016. He is working as the dean of the college of computer engineering and sciences at Prince Sattam bin Abdulaziz University. His research interests include AI security, Applied Cryptography, Access control, and big data analysis and processing.
\end{IEEEbiography}

\EOD

\end{document}